\documentclass[twocolumn,preprintnumbers,aps,prb]{revtex4-1}
\usepackage{graphicx,hyperref}
\usepackage{bm,amsmath,amssymb}
\begin{document}
\title{
One-loop functional renormalization group study for 
the dimensional reduction and its breakdown 
in the long-range random field O($N$) spin model 
near lower critical dimension
}

\author{Yoshinori Sakamoto}
\email{yossi@phys.ge.cst.nihon-u.ac.jp}
\affiliation{
Laboratory of Physics, College of Science and Technology, 
Nihon University, 7-24-1, Narashino-dai, Funabashi-city, Chiba, 
274-8501, Japan
}

\date{\today}
\begin{abstract}
We consider the random-field O($N$) spin model with 
long-range exchange interactions 
which decay with distance $r$ between spins as $r^{-d-\sigma}$
and/or 
random fields which correlate with distance $r$ as $r^{-d+\rho}$ ,  
and reexamine the critical phenomena near the lower critical dimension 
by use of the perturbative functional renormalization group.  
We compute the analytic fixed points in the one-loop beta functions, 
and study their stability. 
We also calculate the critical exponents at the analytical fixed points. 
We show that the analytic fixed point which governs the phase transition 
in the system with the long-range correlations of random fields  
can be destabilized by the nonanalytic perturbation 
in both cases where the exchange interactions between spins are 
short ranged and long ranged. 
For the system with the long-range exchange interactions and uncorrelated random fields, 
we show that the $d\to d-\sigma$ dimensional reduction at the leading order of the $d-2\sigma$ expansion holds only for $N>2(4+3{\sqrt{3}})\simeq 18.3923\cdots$. 
Our investigation into the system 
with the long-range exchange interactions and uncorrelated random fields 
also gives the value of the boundary between critical behaviors 
in systems with long-range and short-range exchange interactions, 
which is identical to that predicted by Sak [Phys. Rev. B {\bf{8}}, 281 (1973)]. 
For the system with the long-range exchange interactions and the long-range correlated random fields,  
we show that the $d\to d-\sigma-\rho$ dimensional reduction does not hold within the present framework, 
as far as $N$ is finite. 
\end{abstract}

\maketitle

\section{Introduction}
The random-field ${\rm O}(N)$ spin model is the model in which 
nonrandom exchange interactions between spins are ferromagnetic 
and external magnetic fields are random. 
To clarify the critical phenomena in this model is one of the fundamental problems 
in the disordered spin system, 
and there are a lot of intensive studies on this \cite{IM,Natt}.  
The $d\to d-\theta$ dimensional reduction gives an important clue 
to clarify the nature of this model. 
The $d\to d-\theta$ dimensional reduction means that 
the effect of random fields reduces the spatial dimension by $\theta$; namely, 
the critical phenomena in $d$-dimensional random-field system 
is equivalent to that in the $(d-\theta)$-dimensional corresponding pure system. 
Here $\theta$ denotes the exponent describing that the flow of the renormalized temperature goes to zero 
under the renormalization-group iteration. 
If the $d\to d-\theta$ dimensional-reduction prediction is correct, 
all critical exponents in the $d$-dimensional random-field system 
should be the same as those in the corresponding pure system in $\theta$ dimensions less.

In the spin system with the short-range ferromagnetic exchange interactions 
and the uncorrelated random fields (SR), 
the $d\to d-2$ dimensional reduction and its breakdown are one of the central issues. 
This conjecture was obtained by the perturbation theory \cite{AIM, Gri,You} 
and the supersymmetry argument \cite{PS}. 
Rigorous proofs have shown that the $d\to d-2$ dimensional-reduction prediction 
is incorrect below four dimensions 
in the case of the random-field Ising model ($N=1$ case) \cite{Imb,BK}.  
The $d\to d-2$ dimensional reduction and its breakdown 
for the random-field ${\rm O}(N)$ spin model above four dimensions 
have been intensively studied. 
Fisher studied the critical phenomena in $4+\epsilon$ dimensions by use of 
the ${\rm O}(N)$ nonlinear-sigma model \cite{Fi}. 
He showed that all possible higher-rank random anisotropies 
which are all relevant operators are generated 
by the perturbative functional renormalization-group iteration 
of the ${\rm O}(N)$ nonlinear-sigma model with only the random field term. 
Then he treated the ${\rm O}(N)$ nonlinear-sigma model 
including the random-field and all the random-anisotropy terms,
and derived the one-loop beta function in $4+\epsilon$ dimensions. 
He showed that there is no singly unstable fixed point 
corresponding to the $d\to d-2$ dimensional reduction at ${\rm O}(\epsilon)$, 
and concluded that the $d\to d-2$ dimensional-reduction prediction is incorrect near four dimensions. 
The one-loop beta function obtained by Fisher 
and the two-loop beta function extended by Le Doussal and Wiese \cite{DW} and Tissier and Tarjus \cite{TT2} 
have been examined carefully \cite{TT2,DW,Fe,SMI,TT3,BTTB}. 
The breakdown of the dimensional reduction is characterized by a nonanalyticity 
which emerges in the first derivative of the function 
including the random-field and all the random-anisotropy terms. 
Namely, the nonanalyticity forms a cusp in the first derivative of the function 
including the random-field and all the random-anisotropy terms, 
which causes the breakdown of the dimensional reduction. 
The singly unstable fixed point corresponding to the $d\to d-2$ dimensional reduction 
exists for $N>18-(49/5)\epsilon$, although it has the weak nonanalyticity 
which does not change the value of the fixed point. 
However, it is unstable with respect to the perturbation with nonanalyticity 
for $N<2(4+3\sqrt{3})-(3(2+3\sqrt{3})/2)\epsilon$. 
Thus, the $d\to d-2$ dimensional reduction holds 
for $N>2(4+3\sqrt{3})-(3(2+3\sqrt{3})/2)\epsilon$, 
and the critical exponents of the connected and the disconnected correlation functions 
$\eta$ and ${\bar{\eta}}$ satisfy ${\bar{\eta}}=\eta$. 
Whereas in $N<2(4+3\sqrt{3})-(3(2+3\sqrt{3})/2)\epsilon$, 
the critical phenomena is governed by the fixed point with the nonanalyticity, 
and thus the $d\to d-2$ dimensional reduction is broken. 
Moreover, a complete theoretical explanation of the $d\to d-2$ dimensional reduction 
and its breakdown has been provided 
through the nonperturbative functional renormalization group \cite{TT1,TT4,TT5,TBT}. 

In the case where the ferromagnetic exchange interactions are short ranged 
and the random fields are correlated over the distance $r$ as $r^{-d+\rho}$ (LRF), 
the $d\to d-2-\rho$ dimensional reduction and its breakdown are still under debate. 
The symbol $\rho$ denotes the exponent 
which characterizes the range of the random-field correlations. 
To study the long-range effect of the random-field correlations in the system, 
we consider the case $0\le\rho<d-2d_{\rm p}$, 
where $d_{\rm p}$ is the lower critical dimension of the corresponding pure system. 
Kardar, McClain, and Taylor performed the renormalization-group calculation near the upper critical dimension $d_u=6+\rho$, 
and concluded that the $d\to d-2-\rho$ dimensional reduction is broken at ${\rm O}(\epsilon^2)$ in $\epsilon=d_u-d$ \cite{KMT}. 
Bray pointed out an error in Kardar, McClain, and Taylor's result 
but their conclusion still holds \cite{Br}.  
Chang and Abrahams carried out the one-loop renormalization-group calculation 
for the ${\rm O}(N)$ nonlinear-sigma model 
near the lower critical dimension $d_l=4+\rho$, 
and showed that the $d\to d-2-\rho$ dimensional reduction is broken 
at ${\rm O}(\epsilon)$ in $\epsilon=d-d_l$ and for $N>3$ \cite{CA1}. 
Fedorenko and K$\ddot{\rm u}$hnel \cite{FK} examined the one-loop beta functions 
of the ${\rm O}(N)$ nonlinear-sigma model 
including not only the uncorrelated and the long-range correlated random fields 
but also all the uncorrelated and the long-range correlated random anisotropies 
which are missed in the work by Chang and Abrahams. 
They showed that the correlation length exponent $\nu$ 
and the phase diagram obtained by Chang and Abrahams are incorrect, 
and the exponents $\eta$ and ${\bar{\eta}}$ are correct only
in a region controlled by the singly unstable fixed point with the weaker nonanalyticity. 

In the case where the long-range ferromagnetic exchange interactions decay with distance $r$ between spins as $r^{-d-\sigma}$
and random fields are uncorrelated (LRE), 
the $d\to d-\sigma$ dimensional reduction and its breakdown are still under debate. 
The symbol $\sigma$ denotes the exponent 
which controls the range of the exchange interactions. 
It should be positive to ensure that the energy density stays finite 
in the thermodynamic limit. 
To study the long-range character of the exchange interactions in the system, 
we consider the case $0<\sigma<2$. 
Young performed the renormalization-group calculation near the upper critical dimension $d_u=3\sigma$, 
and concluded that the $d\to d-\sigma$ dimensional reduction is broken at ${\rm O}(\epsilon^2)$ 
in $\epsilon=d_u-d$ \cite{You}. 
Bray pointed out an error in Young's result 
but the conclusion still holds \cite{Br}.  
Chang and Abrahams carried out the one-loop renormalization-group calculation 
for the ${\rm O}(N)$ nonlinear-sigma model 
near the lower critical dimension $d_l=2\sigma$, 
and showed that the $d\to d-\sigma$ dimensional reduction holds at ${\rm O}(\epsilon)$ in $\epsilon=d-d_l$ and for $N>1$ \cite{CA2}. 
However, the ${\rm O}(N)$ nonlinear-sigma model studied by Chang and Abrahams 
does not contain an infinite number of relevant operators which should be included in the model. 
Recently, Balog, Tarjus, and Tissier studied 
the critical phenomena of a one-dimensional random-field Ising model 
with the long-range exchange interactions and uncorrelated random fields 
by use of the nonperturbative renormalization group, 
and found that there are two distinct regimes characterized by the presence or absence of the nonanalyticity 
in the region of $\frac{1}{3}\le\sigma<\frac{1}{2}$ where the critical exponents take non-classical values  \cite{BTT}. 

In the spin system with the long-range ferromagnetic exchange interactions  
and the long-range correlated random fields (LREF) 
with $0<\sigma<2$ and $0\le\rho<d-2d_{\rm p}$, 
the $d\to d-\sigma-\rho$ dimensional reduction and its breakdown are still under debate. 
Bray used the renormalization-group scaling theory, and showed $\eta=2-\sigma$, $2\eta-{\bar{\eta}}=\rho$, 
and $\theta=\sigma+\rho$. 
Recently, we put $\rho=2-\sigma$, and studied the critical phenomena in the three-dimensional long-range random-field Ising model 
in the region of $1/2<\sigma<1$ by using the nonperturbative functional renormalization group 
combined with the supersymmetric formalism \cite{BTTS}. 
We showed that the $d\to d-2$ dimensional reduction holds for $1/2<\sigma<\sigma_{\rm DR}\approx0.71$,  
and its breakdown is observed in the exponent $\nu$ for $\sigma_{\rm DR}<\sigma<1$. 

In contrast to the case $0<\sigma<2$ 
in which the long-range feature of the exchange interactions is dominant, 
the phase transition for large $\sigma$ belongs to the short-range universality class. 
As the exponent $\sigma$ decreases from large $\sigma$, 
the universality class of the phase transition crosses over 
from the short-range one to the long-range one at a critical value $\sigma_*$. 
In spite of theoretical and numerical studies over forty years, 
the critical behavior in the vicinity of $\sigma=\sigma_*$ is still an ongoing problem. 
There are a lot of studies on this problem in the pure system 
\cite{FMN,Sak,GT,HNH,LB,Pic,BPR,APR,BrePariRi,DTC,HST,BRRZ}.  
In Refs. \onlinecite{FMN,GT} 
it was shown that the exponent $\eta$ changes discontinuously 
from the value in the corresponding short-range system to $2-\sigma$ 
at $\sigma=\sigma_*=2$, 
as $\sigma$ decreases from large $\sigma$. 
In Refs. \onlinecite{Sak,HNH,LB,APR,DTC,HST,BRRZ} 
it was shown that the effect of the long-range exchange interactions is relevant 
for $\sigma<\sigma_*=2-\eta_{\rm sr}$, 
where $\eta_{\rm sr}$ denotes the exponent of the connected correlation function 
in the corresponding short-range system. 
Then the exponent $\eta$ is continuous at $\sigma_*=2-\eta_{\rm sr}$, 
whose value takes $\eta_{\rm sr}$ for $\sigma\ge\sigma_*$, 
and $2-\sigma$ for $\sigma<\sigma_*$. 
Moreover, the presence of a logarithmic correction to the connected correlation function 
at $\sigma_*=2-\eta_{\rm sr}$ was reported in Ref. \onlinecite{BrePariRi}. 
In Refs. \onlinecite{Pic,BPR} 
it was shown that the discontinuity of the exponent $\eta$ 
at $\sigma=2$ does not occur, 
and the value of $\eta$ is interpolated smoothly 
from $\eta_{\rm sr}$ to $2-\sigma$, 
as $\sigma$ decreases from $\sigma=2$. 
In the random-field spin system, 
Bray showed $\sigma_*=2-\eta_{\rm sr}$ 
by using the renormalization-group scaling theory \cite{Br}.  

As stated above, 
the phase transitions in this model are classified into four universality classes (SR, LRF, LRE and LREF), 
according to whether the exchange interactions and/or the random-field correlations in the system 
are short ranged or long ranged. 
However, most studies of the critical phenomena in the random-field ${\rm O}(N)$ spin model 
have been dedicated to the SR case.  
In this paper we consider all four cases. 
We study the critical phenomena near the lower critical dimension 
with the use of the ${\rm O}(N)$ nonlinear-sigma model combined with the replica formalism. 
The model treated in this paper contains 
not only the uncorrelated and the correlated random-field terms 
but also all the uncorrelated and the correlated random-anisotropy terms. 
We employ the perturbative functional renormalization group in order to obtain the one-loop beta functions. 
We examine the properties of the fixed point functions, 
and investigate the stability of the analytic fixed points 
on the basis of the argument by Baczyk, Tarjus, Tissier, and Balog \cite{BTTB}. 
Then we calculate the critical exponents $\eta$, ${\bar{\eta}}$, and $\nu$ 
at each of four analytic fixed points,  
and discuss the critical properties of the system for each universality class. 
We show that the destabilization of the analytic fixed point 
controlling the critical behavior in the system with the long-range correlations 
of random fields can be caused by the perturbation with nonanalyticity 
in both cases where the exchange interactions between spins are 
short ranged and long ranged. 
In the system with LRE, 
we find that the analytic fixed point of O$(\epsilon)$ in $\epsilon=d-d_l$ 
which controls the critical behavior 
is singly unstable not for $N>1$ but for $N>2(4+3\sqrt{3})\simeq 18.3923\cdots$. 
We show that the validity of the $d\to d-\sigma$ dimensional reduction 
at the leading order of the $d-2\sigma$ expansion is confirmed 
only for $N>2(4+3\sqrt{3})\simeq 18.3923\cdots$. 
Moreover, 
by investigating the relation between the critical exponents $\eta$ and ${\bar{\eta}}$, 
we also obtain the critical value $\sigma_*=2-\eta_{\rm sr}$, 
which is the same as that obtained in Refs. \onlinecite{Br,Sak,HNH,LB,APR,DTC,HST,BRRZ}. 
In the system with LREF, 
we find that the analytic fixed point which is singly unstable exists 
under a certain condition. 
However, we show that the $d\to d-\rho-\sigma$ dimensional reduction 
does not hold within the present framework, as far as $N$ is finite. 

The organization of this paper is as follows. 
In Sec. \ref{SRex} 
we study the systems with the short-range exchange interactions, 
namely the SR and LRF cases. 
We perform the one-loop functional renormalization group analysis. 
We show that the analytic fixed point which governs the phase transition 
in the system with LRF can be destabilized by the perturbation with nonanalyticity. 
In Sec. \ref{LRex} 
we study the systems with the long-range exchange interactions, 
namely the LRE and LREF cases. 
We treat the one-loop beta functions, 
and carefully analyze the properties of the analytic fixed points and their stability. 
It is shown that 
the analytic fixed point controlling the critical behavior in the system with LRE 
becomes unstable against the perturbation with nonanalyticity 
for $N<2(4+3\sqrt{3})\simeq18.3923\cdots$, 
which is the same as the case of the system with SR. 
We also show that the destabilization of the analytic fixed point 
which governs the phase transition in the system with LREF 
can occur due to the nonanalytic perturbation. 
As a result, we obtain a certain region in the plane of the parameters $N$ and $\rho$  
where the analytic fixed points are singly unstable. 
In Sec. \ref{LRESRF}, 
we calculate the critical exponents $\eta$, $\bar{\eta}$, and $\nu$ 
at the analytic fixed point 
which controls the critical behavior in the system with LRE. 
We reconsider the validity of the $d\to d-\sigma$ dimensional reduction. 
We also present the result for the critical value $\sigma_*$. 
In Sec. \ref{LRELRF} 
we calculate the critical exponents $\eta$, $\bar{\eta}$, and $\nu$ 
at the analytic fixed point which controls the critical behavior in the system with LREF. 
We show that $d\to d-\sigma-\rho$ dimensional reduction 
breaks down within the present framework, as far as $N$ is finite. 
Sec. \ref{Summary} summarizes our results.

\section{Critical phenomena at zero temperature of long-range correlated random field O($N$) spin model with short-range exchange interactions in $4+\epsilon$ dimensions}\label{SRex}
This section is intended as a reexamination of the critical phenomena 
at zero temperature of the long-range correlated random field O($N$) spin model 
with short-range exchange interactions in $4+\epsilon$ dimensions. 
We discuss the nature of analytic fixed points and their stability. 
And we calculate the critical exponents 
at the analytic fixed point which controls the critical behavior in the system with SR and with LRF. 
\subsection{Model}
Let us consider an $N$-component vector spin system 
where an $N$-component vector spin ${\bm{S}}(x)$ 
with a fixed-length constraint ${\bm{S}}(x)^2=1$ couples to a random field. 
In order to carry out the average over the random field, we use the replica method. 
The critical phenomena  
of the long-range correlated random field O($N$) spin model with the short-range exchange interactions 
near lower critical dimension 
is described by the O($N$) nonlinear-sigma model of the following replica partition function ${\cal Z}$ 
and effective action $\beta H_{\rm rep}$ 
\begin{eqnarray}
{\cal{Z}}
&=&\int\prod_{\alpha=1}^n{\cal{D}}{\mbox{\boldmath$S$}}^{\alpha}
\delta({{\mbox{\boldmath$S$}}^{\alpha}}(x)^2-1)
e^{-\beta H_{\rm rep}},\nonumber\\
\beta H_{\rm rep}
&=&
\frac{a^{2-d}}{2T}\int_x\sum_{\alpha =1}^n 
{\bm{S}}^{\alpha}(x)\cdot(-\partial^2){\bm{S}}^{\alpha}(x)\nonumber\\
&&-\frac{a^{-d}}{2T^2}\int_x\sum_{\alpha,\beta}^n 
R_1
({\bm{S}}^{\alpha}(x)\cdot{\bm{S}}^{\beta}(x))
\nonumber\\
&&-\frac{a^{-d-\rho}}{2T^2}\int_{x,x'}
\sum_{\alpha,\beta}^n 
g(x-x')R_2
({\bm{S}}^{\alpha}(x)\cdot{\bm{S}}^{\beta}(x')),\nonumber\\
\label{actionSR}
\end{eqnarray}
where $a$ is the ultraviolet cutoff, and $\int_x:=\int d^dx$. 
The replica indices denoted 
by Greek indices take values $\alpha,\beta,\ldots = 1,..., n$. 
The first term in the action (\ref{actionSR}) is the kinetic term 
which corresponds to the short-range exchange interactions between spins. 
The parameter $T$ is the dimensionless temperature. 
The function 
$R_i({\bm{S}}^{\alpha}\cdot{\bm{S}}^{\beta})$ ($i=1,2$)
represents the random field and all the random anisotropies, 
and is given by 
\begin{eqnarray}
R_i({\bm{S}}^{\alpha}\cdot{\bm{S}}^{\beta})
=\sum_{r=1}^{\infty}\Delta_{i,r}({\bm{S}}^{\alpha}\cdot{\bm{S}}^{\beta})^r.
\label{RFRAFunctional}
\end{eqnarray}
Here $\Delta_{i,r}$ denotes 
the strength of the random field and the $r$-th rank random anisotropy 
($r=1$ is the random field, 
and $r=2$ is the random second-rank anisotropy). 
The subscript $i=1$ corresponds to 
the uncorrelated random fields and random anisotropies, 
and the subscript $i=2$ corresponds to 
the long-range correlated random fields and random anisotropies with $g(x-x')\sim|x-x'|^{-d+\rho}$. 
The lower critical dimension of this model is $d_l=4+\rho$. 
In the present study, we consider the case of $0\le\rho<\epsilon$. 

\subsection{One-loop beta functions and the zero-temperature fixed points}
To perform the renormalization group transformation, 
we put each replicated vector spin ${\bm{S}}^{\alpha}(x)$ 
as a combination of a slow field ${\bm{n}}_0^{\alpha}(x)$ of the unit length 
and fast fields $\varphi_i^{\alpha}(x)$, $i=1,\ldots,N-1$ such that 
\begin{eqnarray}
{\bm{S}}^{\alpha}(x)
&=&{\bm{n}}_0^{\alpha}(x)
{\sqrt{1-{\bm{\varphi}}^{\alpha}(x)^2}}
+{\bm{\varphi}}^{\alpha}(x)\nonumber\\
&\simeq&{\bm{n}}_0^{\alpha}(x)
-\frac{1}{2}({\bm{\varphi}}^{\alpha}(x)^2){\bm{n}}_0^{\alpha}(x)
+{\bm{\varphi}}^{\alpha}(x),
\label{n_0,e}\\
{\bm{\varphi}}^{\alpha}(x)&=&\sum_{i=1}^{N-1} \varphi_i^{\alpha}(x){\bm{e}}_i^{\alpha}(x),
\label{FastField}
\end{eqnarray}
where the unit vectors ${\mbox{\boldmath$e$}}_i^{\alpha}(x)$ are perpendicular 
to each other and also to the vector 
${\mbox{\boldmath$n$}}_0^{\alpha}(x)$. 
Integrating out the fast fields $\varphi_i^{\alpha}(x)$, 
and calculating the new replicated action 
$\beta H_{\rm rep}'$ up to the second order of the perturbation expansion, 
we get the one-loop beta functions for $T$, $R_1$, and $R_2$, 
which have been obtained by Fedorenko and K$\ddot{\rm u}$hnel \cite{FK}. 
The one-loop beta function for the temperature $T$ is 
\begin{widetext}
\begin{eqnarray}
\partial_tT
=
-(d-2)T+(N-2)T(T+R_1'(1)+R_2'(1)),\label{SR beta func. for T}
\end{eqnarray}
where $\partial_t$ denotes a derivative with respect to $t=\log l$ 
with $l$ being the length-scale parameter which increases toward the infrared direction. 
Here we have rescaled $T$, $R_1$, and $R_2$ by $2/((4\pi)^{d/2}\Gamma(d/2))$. 
We find that $T=0$ is the fixed point, at which the parameter $T$ is irrelevant for $d>2$. 
The one-loop beta functions at $T=0$ for $R_1$ and $R_2$ are 
\begin{eqnarray}
\partial_tR_1(z)
&=&-\epsilon R_1(z)+2(N-2)(R_1'(1)+R_2'(1))R_1(z)-(N-1)z(R_1'(1)+R_2'(1))R_1'(z)\nonumber\\
&&+(1-z^2)(R_1'(1)+R_2'(1))R_1''(z)+\frac{1}{2}(N-2+z^2)(R_1'(z)+R_2'(z))^2\nonumber\\
&&-z(1-z^2)(R_1'(z)+R_2'(z))(R_1''(z)+R_2''(z))+\frac{1}{2}(1-z^2)^2(R_1''(z)+R_2''(z))^2,\label{SREbetafunc1}\\
\partial_tR_2(\zeta)
&=&-(\epsilon-\rho)R_2(\zeta)+2(N-2)(R_1'(1)+R_2'(1))R_2(\zeta)\nonumber\\
&&-(N-1)\zeta(R_1'(1)+R_2'(1))R_2'(\zeta)+(1-\zeta^2)(R_1'(1)+R_2'(1))R_2''(\zeta), 
\label{SREbetafunc2}
\end{eqnarray}
where $z={\bm{n}}_0^{\alpha}(x)\cdot{\bm{n}}_0^{\beta}(x)$, 
$\zeta={\bm{n}}_0^{\alpha}(x)\cdot{\bm{n}}_0^{\beta}(x')$. 
Here we have put $d=4+\epsilon$. 
Practically, the beta functions for the first and second derivatives of $R_1$ and $R_2$ play a central role 
in the critical phenomena at zero temperature near the lower critical dimension. 
The beta functions at $T=0$ for $R_1'$, $R_1''$, $R_2'$, and $R_2''$ in $d=4+\epsilon$ are 
\begin{eqnarray}
\partial_tR_1'(z)
&=&-\epsilon R_1'(z)+(N-3)(R_1'(1)+R_2'(1))R_1'(z)-(N+1)z(R_1'(1)+R_2'(1))R_1''(z)\nonumber\\
&&+(1-z^2)(R_1'(1)+R_2'(1))R_1'''(z)+z(R_1'(z)+R_2'(z))^2\nonumber\\
&&+(N-3+4z^2)(R_1'(z)+R_2'(z))(R_1''(z)+R_2''(z))-z(1-z^2)(R_1'(z)+R_2'(z))(R_1'''(z)+R_2'''(z))\nonumber\\
&&-3z(1-z^2)(R_1''(z)+R_2''(z))^2+(1-z^2)^2(R_1''(z)+R_2''(z))(R_1'''(z)+R_2'''(z)),
\label{SRpartBetaFunc1LRF}\\
\partial_tR_1''(z)
&=&-\epsilon R_1''(z)-4(R_1'(1)+R_2'(1))R_1''(z)-(N+3)z(R_1'(1)+R_2'(1))R_1'''(z)\nonumber\\
&&+(1-z^2)(R_1'(1)+R_2'(1))R_1^{(IV)}(z)+(R_1'(z)+R_2'(z))^2\nonumber\\
&&+10z(R_1'(z)+R_2'(z))(R_1''(z)+R_2''(z))+(N-4+7z^2)(R_1'(z)+R_2'(z))(R_1'''(z)+R_2'''(z))\nonumber\\
&&-z(1-z^2)(R_1'(z)+R_2'(z))(R_1^{(IV)}(z)+R_2^{(IV)}(z))+(N-6+13z^2)(R_1''(z)+R_2''(z))^2\nonumber\\
&&-11z(1-z^2)(R_1''(z)+R_2''(z))(R_1'''(z)+R_2'''(z))+(1-z^2)^2(R_1''(z)+R_2''(z))(R_1^{(IV)}(z)+R_2^{(IV)}(z))\nonumber\\
&&+(1-z^2)^2(R_1'''(z)+R_2'''(z))^2,
\label{SRpartBetaFunc2LRF}
\end{eqnarray}
\begin{eqnarray}
\partial_tR_2'(\zeta)
&=&-(\epsilon-\rho)R_2'(\zeta)+(N-3)(R_1'(1)+R_2'(1))R_2'(\zeta)-(N+1)\zeta(R_1'(1)+R_2'(1))R_2''(\zeta)\nonumber\\
&&+(1-\zeta^2)(R_1'(1)+R_2'(1))R_2'''(\zeta),
\label{LRpartBetaFunc1LRF}\\
\partial_tR_2''(\zeta)
&=&-(\epsilon-\rho)R_2''(\zeta)-4(R_1'(1)+R_2'(1))R_2''(\zeta)-(N+3)\zeta(R_1'(1)+R_2'(1))R_2'''(\zeta)\nonumber\\
&&+(1-\zeta^2)(R_1'(1)+R_2'(1))R_1^{(IV)}(\zeta). 
\label{LRpartBetaFunc2LRF}
\end{eqnarray}

The properties of the fixed point solution $(R_1'(z)^*, R_2'(\zeta)^*)$ are determined 
under the condition that $|R_1'(z)|$ and $|R_2'(\zeta)|$ remain finite during the renormalization group flows.  
We discuss the properties of the fixed point solution $(R_1'(z)^*,R_2'(\zeta)^*)$. 
Eq.(\ref{LRpartBetaFunc1LRF}) is linear in the function $R_2'(\zeta)$, which can be solved analytically. 
Solving the fixed point equation $\partial_tR_2'(\zeta)^*=0$, 
we can find that the fixed point solution $R_2'(\zeta)^*$ is analytic on $\zeta$. 
Next, we assume that the functions $R_1'(z)$ and $R_2'(\zeta)$ take the following form: 
\begin{eqnarray}
R_1'(z)&=&R_1'(1)-R_1''(1)(1-z)+\cdots+a_l(1-z)^{\alpha}+\cdots
\label{initialfunctionLRF1},\\
R_2'(\zeta)&=&R_2'(1)-R_2''(1)(1-\zeta)+\frac{R_2'''(1)}{2}(1-\zeta)^2+\cdots,
\label{initialfunctionLRF2}
\end{eqnarray}
with $\alpha>0$. 
To keep $|R_1'(z)|$ and $|R_2'(\zeta)|$ finite, 
the following condition on the function (\ref{initialfunctionLRF1}) is required; 
\begin{eqnarray}
{\mbox{
$\alpha=\frac{1}{2}\quad$or$\quad\alpha\ge1$
}}.
\label{singular condition for LRF}
\end{eqnarray}
Thus, the fixed point function $R_1(z)^*$ also has the same behavior of $(1-z)^{\alpha^*}$ 
with $\alpha^*=1/2$ or $\alpha^*\ge1$. 
Only in the case of $\alpha^*=1/2$, $R_1''(1)^*$ diverges. 
We use the term ``cuspy'' on a function with $(1-z)^{1/2}$ 
and ``cuspless'' if the first and the second derivatives of a function are finite.

\subsection{Stability of fixed points and critical exponents $\eta$, $\bar{\eta}$ and $\nu$}
The critical exponents $\eta$ and ${\bar{\eta}}$ 
of the connected and disconnected correlation functions 
are expressed by use of $R_1'(1)^*$ and $R_2'(1)^*$ which are the values of $R_1'(1)$ and $R_2'(1)$ 
at the fixed point: 
\begin{eqnarray}
&&\eta=R_1'(1)^*+R_2'(1)^*,\\
&&{\bar{\eta}}=(N-1)(R_1'(1)^*+R_2'(1)^*)-\epsilon.
\end{eqnarray}
The critical exponent $\nu$ of the correlation length 
is given by the inverse of the maximal eigenvalue of the scaling matrix at the fixed point. 
Then, we find the fixed points 
by solving $\partial_tR_1'(1)^*=0$, $\partial_tR_1''(1)^*=0$, $\partial_tR_2'(1)^*=0$, and $\partial_tR_2''(1)^*=0$, 
study their stability, 
and calculate the critical exponents $\eta$, $\bar{\eta}$, and $\nu$ in the following. 

The fixed points are 
\begin{eqnarray}
(R_1'(1)^*,R_2'(1)^*,R_{1+}''(1)^*,R_2''(1)^*)
&=&\biggl(
\frac{\epsilon}{N-2},0,
\frac{\epsilon[N-8+\sqrt{(N-2)(N-18)}]}{2(N+7)(N-2)},0
\biggr),\label{SR FP+}\\
(R_1'(1)^*,R_2'(1)^*,R_{1-}''(1)^*,R_2''(1)^*)
&=&\biggl(
\frac{\epsilon}{N-2},0,
\frac{\epsilon[N-8-\sqrt{(N-2)(N-18)}]}{2(N+7)(N-2)},0
\biggr),\label{SR FP-}\\
(R_1'(1)^*,R_2'(1)^*,R_{1+}''(1)^*,R_2''(1)^*)
&=&\biggl(
\frac{(\epsilon-\rho)^2}{(N-3)^2\rho},
\frac{(\epsilon-\rho)^2}{(N-3)^2\rho}\{(N-3){\hat{\epsilon}}-(N-2)\},
\nonumber\\
&&\frac{(\epsilon-\rho)
[(N-3)\hat{\epsilon}-6
+\sqrt{
\{(N-3)\hat{\epsilon}-6\}^2-4(N+7)
}]
}{2(N+7)(N-3)},
0
\biggr),\label{LRF FP+}\\
(R_1'(1)^*,R_2'(1)^*,R_{1-}''(1)^*,R_2''(1)^*)
&=&\biggl(
\frac{(\epsilon-\rho)^2}{(N-3)^2\rho},
\frac{(\epsilon-\rho)^2}{(N-3)^2\rho}\{(N-3){\hat{\epsilon}}-(N-2)\},
\nonumber\\
&&\frac{(\epsilon-\rho)
[(N-3)\hat{\epsilon}-6
-\sqrt{
\{(N-3)\hat{\epsilon}-6\}^2-4(N+7)
}]
}{2(N+7)(N-3)},
0
\biggr).\label{LRF FP-}
\end{eqnarray}
\end{widetext}
Here we have introduced the reduced variable ${\hat{\epsilon}}$: 
\begin{eqnarray}
\hat{\epsilon}=\frac{\epsilon}{\epsilon-\rho}. 
\end{eqnarray}
The stability of the cuspless fixed points with respect to the cuspless perturbation 
can be investigated by calculating eigenvalues of the $4\times 4$ scaling matrix whose elements are 
the first derivatives of the beta functions 
$\partial_tR_1'(1)$, $\partial_tR_1''(1)$, $\partial_tR_2'(1)$, and $\partial_tR_2''(1)$
at the cuspless fixed points. 

The cuspless fixed points (\ref{SR FP+}) and (\ref{SR FP-}) exist for $N\ge18$. 
The eigenvalues $\lambda_1,\ldots,\lambda_4$ of the scaling matrix at the cuspless fixed points 
(\ref{SR FP+}) and (\ref{SR FP-}) are given by 
\begin{eqnarray}
\lambda_1&=&\epsilon,\\
\lambda_2&=&(\epsilon-\rho)\frac{(N-3){\hat{\epsilon}}-(N-2)}{N-2},
\label{SRev}\\
\lambda_3^{\pm}&=&\pm\epsilon\sqrt{\frac{N-18}{N-2}},\label{SRev2}\\
\lambda_4&=&-\biggl(\epsilon-\rho+\frac{4\epsilon}{N-2}\biggr),
\end{eqnarray}
Thus, the cuspless fixed point (\ref{SR FP+}) is multiply unstable. 
If $1<{\hat{\epsilon}}<(N-2)/(N-3)$, namely $\lambda_2<0$, 
the cuspless fixed point (\ref{SR FP-}) is singly unstable. 
Due to $R_2'(1)^*=0$, 
the long-range correlations of random fields and random anisotropies 
are irrelevant, 
and thus the cuspless fixed point (\ref{SR FP-}) governs the phase transition 
in the system with SR. 
The critical exponents $\eta_{\rm SR}$ of the connected correlation function 
and ${\bar{\eta}}_{\rm SR}$ of the disconnected correlation function 
at the cuspless fixed point (\ref{SR FP-}) are 
\begin{eqnarray}
\eta_{\rm SR}&=&\frac{\epsilon}{N-2},\label{DR eta_SR}\\
{\bar{\eta}}_{\rm SR}&=&\frac{\epsilon}{N-2}.\label{DR bareta_SR}
\end{eqnarray}
And the critical exponent $\nu_{\rm SR}$ 
which characterizes the divergence of the correlation length 
in the vicinity of transition is 
\begin{eqnarray}
\nu_{\rm SR}=\frac{1}{\epsilon}. 
\end{eqnarray}
Whereas in $N\le18$ 
the cuspless fixed points (\ref{SR FP+}) and (\ref{SR FP-}) merge 
and annihilate, and thus 
the beta functions have no cuspless fixed point of O($\epsilon$). 

The cuspless fixed points (\ref{LRF FP+}) and (\ref{LRF FP-}) exist 
for $N>3$ and 
\begin{eqnarray}
\hat{\epsilon}
\ge\hat{\epsilon}_{\rm cuspless}
=\frac{6+2\sqrt{N+7}}{N-3}.
\end{eqnarray}
The eigenvalues $\lambda_1,\ldots,\lambda_4$ of the scaling matrix at the fixed points 
(\ref{LRF FP+}) and (\ref{LRF FP-}) are given by 
\begin{eqnarray}
\lambda_1&=&(\epsilon-\rho)\frac{N-2}{N-3}-\frac{\epsilon}{2}
\nonumber\\
&&+\frac{\epsilon}{2}
\sqrt{1+\frac{4[N-2-{\hat{\epsilon}}(N-3)]}{{\hat{\epsilon}}^2(N-3)^2}},\\
\lambda_2&=&(\epsilon-\rho)\frac{N-2}{N-3}-\frac{\epsilon}{2}
\nonumber\\
&&-\frac{\epsilon}{2}
\sqrt{1+\frac{4[N-2-{\hat{\epsilon}}(N-3)]}{{\hat{\epsilon}}^2(N-3)^2}},
\label{LRFev}\\
\lambda_3^{\pm}&=&
\pm(\epsilon-\rho)\frac{\sqrt{\{(N-3)\hat{\epsilon}-6\}^2-4(N+7)}}{N-3},
\label{LRFev2}\\
\lambda_4&=&-(\epsilon-\rho)\frac{N+1}{N-3},
\end{eqnarray}
Thus, the cuspless fixed point (\ref{LRF FP+}) is multiply unstable. 
If ${\hat{\epsilon}}>(N-2)/(N-3)$, namely $\lambda_2<0$, 
the cuspless fixed point (\ref{LRF FP-}) is singly unstable. 
Due to $R_2'(1)^*>0$, 
the effect of the long-range correlation of random fields and random anisotropies appears, 
and then the cuspless fixed point (\ref{LRF FP-}) governs the phase transition 
in the system with LRF. 
Thus, the critical exponents $\eta_{\rm LRF}$ and ${\bar{\eta}}_{\rm LRF}$ 
at the cuspless fixed point (\ref{LRF FP-}) are 
\begin{eqnarray}
\eta_{\rm LRF}&=&\frac{\epsilon-\rho}{N-3},\\
{\bar{\eta}}_{\rm LRF}&=&\frac{2\epsilon-(N-1)\rho}{N-3}. 
\end{eqnarray}
These exponents satisfy the Schwartz-Soffer inequality 
${\bar{\eta}}_{\rm LRF}\le 2\eta_{\rm LRF}$ \cite{SS},  
and saturate the generalized Schwartz-Soffer inequality 
${\bar{\eta}}_{\rm LRF}\le 2\eta_{\rm LRF}-\rho$ \cite{VS1}.  
And the inverse of the exponent $\nu_{\rm LRF}$ is 
\begin{eqnarray}
\nu_{\rm LRF}^{-1}
&=&\frac{(N-2)(\epsilon-\rho)}{N-3}
\biggl[
1-\frac{(N-3){\hat{\epsilon}}}{2(N-2)}
\nonumber\\
&&+\frac{(N-3){\hat{\epsilon}}}{2(N-2)}
\sqrt{1+\frac{4[N-2-{\hat{\epsilon}}(N-3)]}{{\hat{\epsilon}}^2(N-3)^2}}
\biggr].
\end{eqnarray}
Whereas in $\hat{\epsilon}\le(6+2\sqrt{N+7})/(N-3)$ 
the cuspless fixed points (\ref{LRF FP+}) and (\ref{LRF FP-}) merge and annihilate, 
and thus the beta functions have no cuspless fixed point of O($\epsilon$). 

\begin{figure}
\begin{center}
\includegraphics[width=80mm]{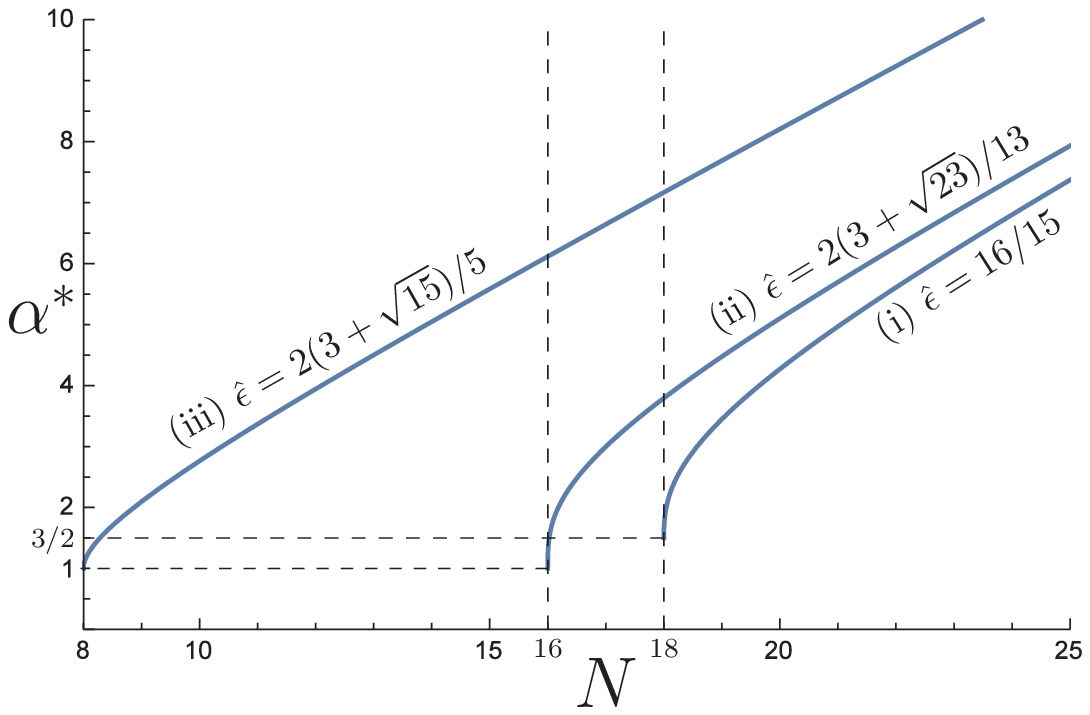}
\end{center}
\caption{
Exponent $\alpha^*=\alpha(N,{\hat{\epsilon}})^*$ characterizing the nonanalyticity $(1-z)^{\alpha^*}$ 
of the LRF TT FP (\ref{LRF FP-}) 
for three values of ${\hat{\epsilon}}$. 
The ordinate is $\alpha^*$, and the abscissa is $N$. 
(i) For ${\hat{\epsilon}}=16/15$, the value $N$ of the lower boundary 
above which the LRF TT FP are singly unstable with respect to the cuspless perturbation is $N=18$. 
$\alpha(N\searrow18,{\hat{\epsilon}})^*=3/2$. 
(ii) For ${\hat{\epsilon}}=2(3+\sqrt{23})/13$, $N=16$. 
$\alpha(N\searrow16,{\hat{\epsilon}})^*=1$. 
(iii) For ${\hat{\epsilon}}=2(3+\sqrt{15})/5$, $N=8$. 
$\alpha(N\searrow8,{\hat{\epsilon}})^*=1$. 
}
\label{alpha vs N LRF}
\end{figure}

As Tissier and Tarjus (TT) and co-workers argued in Ref. \onlinecite{TT1,TT2,TT3,TT4,BTTB}, 
the cuspless fixed points (\ref{SR FP-}) and (\ref{LRF FP-}) have weaker nonanalyticities 
$(1-z)^{\alpha^*}$ with a noninteger $\alpha^*\ge1$. 
The weaker nonanalyticity is called ``subcusp''. 
We refer to the cuspless fixed points (\ref{SR FP-}) and (\ref{LRF FP-}) 
as ``SR TT FP'' and ``LRF TT FP'', respectively. 
The weaker nonanalyticity does not alter the flow equations 
for $R_1'(1)$ and $R_2'(1)$. 
The power $\alpha^*$ is obtained as follows. 
Calculating the flow of $a_l$ in Eq. (\ref{initialfunctionLRF1}), 
we have 
\begin{eqnarray}
&&\partial_ta_l=a_l\Lambda_{\alpha+1}(R_1'(1)^*,R_2'(1)^*,R_1''(1)^*,R_2''(1)^*),\\
&&\Lambda_{\alpha+1}(R_1'(1)^*,R_2'(1)^*,R_1''(1)^*,R_2''(1)^*)\nonumber\\
&&=2[R_1'(1)^*+R_2'(1)^*+3(R_1''(1)^*+R_2''(1)^*)]\alpha^2\nonumber\\
&&-[(N-5)(R_1'(1)^*+R_2'(1)^*)\nonumber\\
&&\quad-(N+7)(R_1''(1)^*+R_2''(1)^*)]\alpha\nonumber\\
&&+(N-1)(R_1'(1)^*+R_2'(1)^*)\nonumber\\
&&+(N+1)(R_1''(1)^*+R_2''(1)^*)-\epsilon.
\label{Eigen for a_l of LRF}
\end{eqnarray}
The power $\alpha^*$ is determined from 
\begin{eqnarray}
\Lambda_{\alpha^*+1}(R_1'(1)^*,R_2'(1)^*,R_1''(1)^*,R_2''(1)^*)=0.
\label{subcusp eq. of LRF}
\end{eqnarray}
Substituting the SR TT FP (\ref{SR FP-}) 
and the LRF TT FP (\ref{LRF FP-}) 
into the above equation, we have explicit expressions for $\alpha^*$, respectively. 
Here we treat only the LRF case (see Ref. \onlinecite{TT3} for the SR case). 
From Eqs.(\ref{Eigen for a_l of LRF}) and (\ref{subcusp eq. of LRF}), 
we obtain the following quadratic equation for $\alpha^*$: 
\begin{eqnarray}
&&\biggl(
2+\frac{3}{N+7}
[(N-3){\hat{\epsilon}}-6\nonumber\\
&&-\sqrt{\{(N-3){\hat{\epsilon}}-6\}^2-4(N+7)}]
\biggr){\alpha^*}^2\nonumber\\
&&-\biggl(
N-5-\frac{1}{2}
[(N-3){\hat{\epsilon}}-6\nonumber\\
&&-\sqrt{\{(N-3){\hat{\epsilon}}-6\}^2-4(N+7)}]
\biggr)\alpha^*\nonumber\\
&&+N-1-(N-3){\hat{\epsilon}}\nonumber\\
&&+\frac{N+1}{2(N+7)}
[(N-3){\hat{\epsilon}}-6\nonumber\\
&&-\sqrt{\{(N-3){\hat{\epsilon}}-6\}^2-4(N+7)}]=0.
\end{eqnarray}
Solving the above quadratic equation, we obtain the solution $\alpha^*$ 
as a function of $N$ and ${\hat{\epsilon}}$. 
It goes to $N/2+{\rm O}(1)$ at large $N$. 
The graphs of $\alpha^*=\alpha(N,{\hat{\epsilon}})^*$ for some values of ${\hat{\epsilon}}$ 
are depicted in Fig. \ref{alpha vs N LRF}. 

We proceed to investigate the stability of the cuspless fixed points 
with respect to the cuspy perturbation, 
following the work by Baczyk, Tarjus, Tissier, and Balog \cite{BTTB}. 
The eigenvalue $\lambda$ relating to the cuspy deformation 
from the cuspless fixed points is given by 
\begin{eqnarray}
\lambda
=\Lambda_{3/2}(R_1'(1)^*,R_2'(1)^*,R_1''(1)^*,R_2''(1)^*)
\label{ev wrt cuspy pert of LRF}
\end{eqnarray}
Substituting the SR TT FP (\ref{SR FP-}) 
and the LRF TT FP (\ref{LRF FP-}) 
into the above equation, we have explicit expressions for $\lambda$, respectively. 
The eigenvalues $\lambda_{\rm SR}$ for the system with SR 
and $\lambda_{\rm LRF}$ for the system with LRF are as follows: 
\begin{eqnarray} 
\lambda_{\rm SR}
&=&-\frac{\epsilon}{4(N+7)}\biggl[3(N+4)\sqrt{\frac{N-18}{N-2}}-N+8\biggr],\\
\lambda_{\rm LRF}
&=&-\frac{\epsilon-\rho}{4(N-3)(N+7)}\nonumber\\
&&\times[3(N+4)\sqrt{\{(N-3){\hat{\epsilon}}-6\}^2-4(N+7)}\nonumber\\
&&+(N+16)\{(N-3){\hat{\epsilon}}-6\}-2(N+7)(N-8)].\nonumber\\
\end{eqnarray}

In the case of the system with SR, we find that, 
below $N=N_{\rm cusp}=2(4+3\sqrt{3})\simeq18.3923\ldots$, 
the eigenvalue $\lambda_{\rm SR}$ takes a positive value. 
Thus the cuspy perturbation becomes relevant for $N<N_{\rm cusp}$, 
where the SR TT FP (\ref{SR FP-}) is multiply unstable with respect to the cuspy perturbation. 
Whereas it remains singly unstable with respect to the perturbation with and without the cuspy behavior 
for $N>N_{\rm cusp}$. 
As shown in Ref. \onlinecite{BTTB}, 
there exists a singly unstable cuspy fixed point below $N=N_{\rm cusp}$. 
As $N$ decreases from sufficiently large $N$, 
the fixed point which governs the phase transition in the system continuously changes 
from the SR TT FP to the singly unstable cuspy fixed point at $N=N_{\rm cusp}$ 
before $N$ reaches to $N=18$. 
Accordingly, the values of the critical exponents $\eta_{\rm SR}$ and ${\bar{\eta}}_{\rm SR}$  
deviate from the dimensional-reduction results (\ref{DR eta_SR}) and (\ref{DR bareta_SR}) 
below $N=N_{\rm cusp}$. 

In the case of the system with LRF, the eigenvalue $\lambda_{\rm LRF}$ takes a positive value 
below 
\begin{eqnarray}
&&{\hat{\epsilon}}={\hat{\epsilon}}_{\rm cusp}\nonumber\\
&&=\frac{3(N+4)\sqrt{N^2-8N+48}-(N+4)(N-20)}{4(N-2)(N-3)}.
\end{eqnarray}
Since ${\hat{\epsilon}}_{\rm cusp}\ge{\hat{\epsilon}}_{\rm cuspless}$ 
for $4(1+\sqrt{7})\le N \le 2(4+3\sqrt{3})$, 
the LRF TT FP (\ref{LRF FP-}) is destabilized by the cuspy perturbation 
for $4(1+\sqrt{7})\le N \le 2(4+3\sqrt{3})$ and ${\hat{\epsilon}}<{\hat{\epsilon}}_{\rm cusp}$. 
Even in this case, a singly unstable cuspy fixed point which governs the phase transition in the system 
is considered to exist 
for $4(1+\sqrt{7})\le N \le 2(4+3\sqrt{3})$ and ${\hat{\epsilon}}<{\hat{\epsilon}}_{\rm cusp}$. 

Finally, we calculate the eigenfunction 
which belongs to the eigenvalue (\ref{ev wrt cuspy pert of LRF}). 
Solving the eigenvalue equation, 
we obtain two solutions. 
One takes the form of $(1-z)^{\alpha_-(\lambda)}$ with $\alpha_-(\lambda)=1/2$ when $z\to 1$, 
and the other takes the form of $(1-z)^{\alpha_+(\lambda)}$. 
Both solutions individually diverge in $z=-1$. 
The physical eigenfunction is represented as a linear combination 
of two solutions of the eigenvalue equation, 
in which the coefficients should be chosen to eliminate the singularities at $z=-1$. 
The power $\alpha_+(\lambda)$ of the function $(1-z)^{\alpha_+(\lambda)}$ 
can be obtained by imposing 
\begin{eqnarray}
&&\Lambda_{\alpha_++1}(R_1'(1)^*,R_2'(1)^*,R_1''(1)^*,R_2''(1)^*)\nonumber\\
&&=\Lambda_{3/2}(R_1'(1)^*,R_2'(1)^*,R_1''(1)^*,R_2''(1)^*).
\label{power of subcusp in cuspy deform for LRF}
\end{eqnarray}

In the case of the system with SR, 
substituting the SR TT FP (\ref{SR FP-}) into Eq. (\ref{power of subcusp in cuspy deform for LRF}), 
we have 
\begin{eqnarray}
\alpha_+(\lambda_{\rm SR})=\frac{1}{4}(N-10+\sqrt{(N-2)(N-18)}). 
\end{eqnarray}
For $N\ge 18$, $\alpha_+(\lambda_{\rm SR})$ takes $\alpha_+(\lambda_{\rm SR})\ge 2$. 

In the case of the system with LRF, 
substituting the LRF TT FP (\ref{LRF FP-}) into Eq. (\ref{power of subcusp in cuspy deform for LRF}), 
we have 
\begin{eqnarray}
\alpha_+(\lambda_{\rm LRF})
&=&
\frac{(N-14)(N-3){\hat{\epsilon}}+(N-2)(N-6)}{2[3(N-3){\hat{\epsilon}}+N-2]}\nonumber\\
&&+\frac{\sqrt{\{(N-3){\hat{\epsilon}}-6\}^2-4(N+7)}}{3(N-3){\hat{\epsilon}}+N-2}. 
\end{eqnarray} 
The power $\alpha_+(\lambda_{\rm LRF})$ takes 
$\alpha_+(\lambda_{\rm LRF})\ge 1+\sqrt{3}$ 
for $N\ge N_{\rm cusp}$ and ${\hat{\epsilon}}\ge(N-2)/(N-3)$, 
and $\alpha_+(\lambda_{\rm LRF})\ge 1$ 
for $4(1+\sqrt{7})\le N<N_{\rm cusp}$ and ${\hat{\epsilon}}\ge{\hat{\epsilon}}_{\rm cuspless}$. 
However, we should note that, 
for $N<4(1+\sqrt{7})$ and in the region of 
\begin{eqnarray}
{\hat{\epsilon}}_{\rm cuspless}\le{\hat{\epsilon}}<\frac{(N-2)(N^2+32)}{(N-3)(N-8)(N+16)}, 
\label{alphaLRF<1}
\end{eqnarray}
$\alpha_+(\lambda_{\rm LRF})<1$, 
which is in contradiction with the condition (\ref{singular condition for LRF}). 
Thus, for $N<4(1+\sqrt{7})$ and in the region (\ref{alphaLRF<1}), 
the cuspy deformation from the LRF TT FP (\ref{LRF FP-}) is unphysical. 
Then, the destabilization of the LRF TT FP (\ref{LRF FP-}) by the cuspy perturbation does not occur 
for ${\hat{\epsilon}}>2(23+19\sqrt{7})/111$. 

\begin{figure}
\begin{center}
\includegraphics[width=80mm]{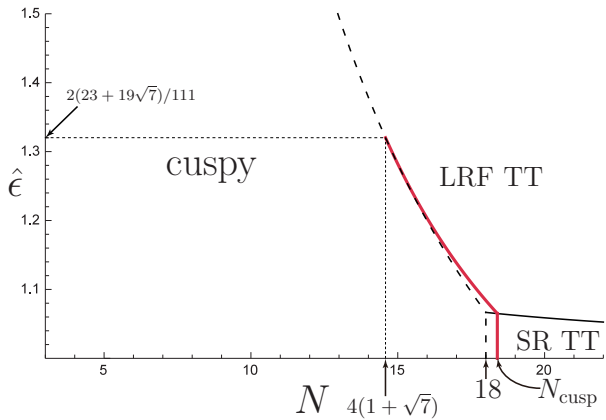}
\end{center}
\caption{
The regions where various FPs are singly unstable. 
The ordinate is ${\hat{\epsilon}}(=\epsilon/(\epsilon-\rho))$, 
and the abscissa is $N$. 
The broken line (black) denotes the lower boundary 
above which the SR TT and the LRF TT FPs are singly unstable against the cuspless perturbation. 
The border line (black line) between the SR TT and LRF TT FPs 
is given by ${\hat{\epsilon}}=(N-2)/(N-3)$. 
The solid line (red line) denotes the lower boundary 
above which the SR TT and LRF TT FPs are singly unstable against the cuspy perturbation. 
}
\label{phasediagramLRF}
\end{figure}

The regions where the various fixed points are singly unstable are depicted 
in Fig. \ref{phasediagramLRF}. 
Outside the areas where the SR TT and the LRF TT FPs are singly unstable, 
the cuspy fixed point is considered to control the critical behavior in the system. 
Particularly, in the region of $1\le{\hat{\epsilon}}<2(23+19\sqrt{7})/111\simeq1.32$, 
the destabilization of the SR TT and the LRF TT FPs by the cuspy perturbation is caused 
at $N_{\rm cusp}$ for the SR TT FP, and at ${\hat{\epsilon}}_{\rm cusp}$ for the LRF TT FP, 
respectively. 

\section{Fixed points and their stability in the renormalization group 
of long-range correlated random field O($N$) spin model 
with long-range exchange interactions in $2\sigma+\rho+\epsilon$ dimensions}\label{LRex}
We now study the critical phenomena at zero temperature 
of the long-range correlated random field O($N$) spin model 
with the long-range exchange interactions in $2\sigma+\rho+\epsilon$ dimensions 
by use of the renormalization group. 
The critical phenomena at zero temperature 
of the long-range correlated random field O($N$) spin model 
with the long-range exchange interactions near lower critical dimension 
is described by the O$(N)$ nonlinear-sigma model. 
In this section 
we investigate the fixed points and their stability of the one-loop beta functions 
in the O$(N)$ nonlinear-sigma model. 
The critical phenomena are carefully discussed in the subsequent sections. 
\subsection{Model}
We start from the O($N$) nonlinear-sigma model with the replica effective action  
\begin{eqnarray}
\beta H_{\rm rep}
&=&
\frac{a^{\sigma-d}}{2T}\int_x\sum_{\alpha =1}^n 
{\bm{S}}^{\alpha}(x)\cdot(-\partial^2)^{\sigma/2}{\bm{S}}^{\alpha}(x)\nonumber\\
&&-\frac{a^{-d}}{2T^2}\int_x\sum_{\alpha,\beta}^n 
R_1
({\bm{S}}^{\alpha}(x)\cdot{\bm{S}}^{\beta}(x))
\nonumber\\
&&-\frac{a^{-d-\rho}}{2T^2}\int_{x,x'}
\sum_{\alpha,\beta}^n 
g(x-x')R_2
({\bm{S}}^{\alpha}(x)\cdot{\bm{S}}^{\beta}(x')).\nonumber\\
\label{actionLR}
\end{eqnarray}
The first term in the action (\ref{actionLR}) is the kinetic term 
which corresponds to the long-range exchange interactions between spins. 
The operator $(-\partial^2)^{\sigma/2}$ denotes the fractional Laplacian in the Euclidean space.  
In the present study we consider the case of $0<\sigma<2$. 
The parameter $T$ denotes the dimensionless temperature. 
The function 
$R_i(z)$ ($i=1,2$)
represents the random field and all the random anisotropies, 
which is defined by Eq.(\ref{RFRAFunctional}).  
The lower critical dimension of this model is $d_l=2\sigma+\rho$. 
In the present study, we consider the case of $\rho\ge0$.

\subsection{One-loop beta functions}
To carry out the renormalization group transformation, 
it is convenient to use the momentum representation. 
The fractional Laplacian $(-\partial^2)^{\sigma/2}$ is written by its Fourier transformation: 
\begin{eqnarray}
(-\partial^2)^{\sigma/2}f(x)=\int_kk^{\sigma}{\tilde{f}}(k)e^{ikx},
\end{eqnarray}
\begin{widetext}
\noindent
where 
$kx=k^{(1)}x^{(1)}+\cdots+k^{(d)}x^{(d)}$, 
$k^{\sigma}=({k^{(1)}}^2+\cdots+{k^{(d)}}^2)^{\sigma/2}$, 
and $\int_k\equiv\int\frac{d^dk}{(2\pi)^d}$. 
The correlation of the random fields $g(x-x')$ is written as 
\begin{eqnarray}
g(x-x')\sim\frac{1}{|x-x'|^{d-\rho}}
=\int_kk^{-\rho}e^{ik(x-x')},
\end{eqnarray}
in the momentum representation. 
The $N$-component replicated vector spin ${\mbox{\boldmath$S$}}^{\alpha}(x)$ 
of the magnetization (\ref{n_0,e}) 
is rewritten in the momentum representation as follows: 
\begin{eqnarray}
{\mbox{\boldmath$S$}}^{\alpha}(x)
&\simeq&{\mbox{\boldmath$n$}}_0^{\alpha}(x)
-\frac{1}{2}({{\mbox{\boldmath$\varphi$}}^{\alpha}(x)}^2)
{\mbox{\boldmath$n$}}_0^{\alpha}(x)
+{\mbox{\boldmath$\varphi$}}^{\alpha}(x)\nonumber\\
&=&\int_k{\tilde{{\mbox{\boldmath$n$}}}}_0^{\alpha}(k)e^{ikx}
-\frac{1}{2}\int_{k,k_1,\cdots,k_4}
\biggl(\sum_{i,j}^{N-1}
{\tilde{\varphi}}_i^{\alpha}(k_1){\tilde{\varphi}}_j^{\alpha}(k_2)
{\tilde{{\mbox{\boldmath$e$}}}}_i^{\alpha}(k_3)\cdot
{\tilde{{\mbox{\boldmath$e$}}}}_j^{\alpha}(k_4)
\biggr){\tilde{{\mbox{\boldmath$n$}}}}_0^{\alpha}(k)
e^{i(k_1+\cdots+k_4+k)x}\nonumber\\
&&+\int_{k_1,k_2}
\biggl(\sum_{i=1}^{N-1}
{\tilde{\varphi}}_i^{\alpha}(k_1)
{\tilde{{\mbox{\boldmath$e$}}}}_i^{\alpha}(k_2)
\biggr)
e^{i(k_1+k_2)x}.
\label{n_0,e,FT}
\end{eqnarray}
We integrate out the fast fields ${\tilde{\varphi}}_i^{\alpha}(k)$, 
and calculate the new replicated action 
$\beta H_{\rm rep}'$ up to the second order of the perturbation expansion. 
After rewriting $\beta H_{\rm rep}'$ in the coordinate representation again, 
we can then obtain the one-loop beta functions for $T$, $R_1$, and $R_2$. 
The one-loop beta function for the temperature $T$ is 
\begin{eqnarray}
\partial_tT=
-(d-\sigma)T+(N-1)T(T+R_1'(1)+R_2'(1)).\label{beta func. for T}
\end{eqnarray}
Here we have rescaled $T$, $R_1$, and $R_2$ by $2/((4\pi)^{d/2}\Gamma(d/2))$. 
For $d>\sigma$, 
we find that $T=0$ is the fixed point, 
at which the parameter $T$ is irrelevant. 
The one-loop beta functions at $T=0$ for $R_1$ and $R_2$ are
\begin{eqnarray}
\partial_tR_1(z)
&=&-(\epsilon+\rho)R_1(z)+2(N-1)(R_1'(1)+R_2'(1))R_1(z)-(N-1)z(R_1'(1)+R_2'(1))R_1'(z)\nonumber\\
&&+(1-z^2)(R_1'(1)+R_2'(1))R_1''(z)+\frac{1}{2}(N-2+z^2)(R_1'(z)+R_2'(z))^2\nonumber\\
&&-z(1-z^2)(R_1'(z)+R_2'(z))(R_1''(z)+R_2''(z))+\frac{1}{2}(1-z^2)^2(R_1''(z)+R_2''(z))^2,
\label{LREbetafunc1}\\
\partial_tR_2(\zeta)
&=&-\epsilon R_2(\zeta)+2(N-1)(R_1'(1)+R_2'(1))R_2(\zeta)-(N-1)\zeta(R_1'(1)+R_2'(1))R_2'(\zeta)\nonumber\\
&&+(1-\zeta^2)(R_1'(1)+R_2'(1))R_2''(\zeta).
\label{LREbetafunc2}
\end{eqnarray}
Here we have put $d=2\sigma+\rho+\epsilon$. 
To study the fixed points and their stability, 
we consider the beta functions for their derivative. 
Differentiating Eqs.(\ref{LREbetafunc1}) and (\ref{LREbetafunc2}) 
with respect to $z$ and $\zeta$, respectively, 
we obtain the one-loop beta functions for $R_1'(z)$, $R_1''(z)$, $R_2'(\zeta)$, 
and $R_2''(\zeta)$; 
\begin{eqnarray}
\partial_tR_1'(z)
&=&-(\epsilon+\rho) R_1'(z)+(N-1)(R_1'(1)+R_2'(1))R_1'(z)-(N+1)z(R_1'(1)+R_2'(1))R_1''(z)\nonumber\\
&&+(1-z^2)(R_1'(1)+R_2'(1))R_1'''(z)+z(R_1'(z)+R_2'(z))^2\nonumber\\
&&+(N-3+4z^2)(R_1'(z)+R_2'(z))(R_1''(z)+R_2''(z))-z(1-z^2)(R_1'(z)+R_2'(z))(R_1'''(z)+R_2'''(z))\nonumber\\
&&-3z(1-z^2)(R_1''(z)+R_2''(z))^2+(1-z^2)^2(R_1''(z)+R_2''(z))(R_1'''(z)+R_2'''(z)),
\label{SRpartBetaFunc1LREF}\\
\partial_tR_1''(z)
&=&-(\epsilon+\rho) R_1''(z)-2(R_1'(1)+R_2'(1))R_1''(z)-(N+3)z(R_1'(1)+R_2'(1))R_1'''(z)\nonumber\\
&&+(1-z^2)(R_1'(1)+R_2'(1))R_1^{(IV)}(z)+(R_1'(z)+R_2'(z))^2\nonumber\\
&&+10z(R_1'(z)+R_2'(z))(R_1''(z)+R_2''(z))+(N-4+7z^2)(R_1'(z)+R_2'(z))(R_1'''(z)+R_2'''(z))\nonumber\\
&&-z(1-z^2)(R_1'(z)+R_2'(z))(R_1^{(IV)}(z)+R_2^{(IV)}(z))+(N-6+13z^2)(R_1''(z)+R_2''(z))^2\nonumber\\
&&-11z(1-z^2)(R_1''(z)+R_2''(z))(R_1'''(z)+R_2'''(z))+(1-z^2)^2(R_1''(z)+R_2''(z))(R_1^{(IV)}(z)+R_2^{(IV)}(z))\nonumber\\
&&+(1-z^2)^2(R_1'''(z)+R_2'''(z))^2,
\label{SRpartBetaFunc2LREF}\\
\partial_tR_2'(\zeta)
&=&-\epsilon R_2'(\zeta)+(N-1)(R_1'(1)+R_2'(1))R_2'(\zeta)-(N+1)\zeta(R_1'(1)+R_2'(1))R_2''(\zeta)\nonumber\\
&&+(1-\zeta^2)(R_1'(1)+R_2'(1))R_2'''(\zeta),
\label{LRpartBetaFunc1LREF}\\
\partial_tR_2''(\zeta)
&=&-\epsilon R_2''(z)-2(R_1'(1)+R_2'(1))R_2''(\zeta)-(N+3)\zeta(R_1'(1)+R_2'(1))R_2'''(\zeta)\nonumber\\
&&+(1-\zeta^2)(R_1'(1)+R_2'(1))R_1^{(IV)}(\zeta).
\label{LRpartBetaFunc2LREF}
\end{eqnarray}

We discuss the properties of the fixed point solution $(R_1'(z)^*,R_2'(\zeta)^*)$. 
First, we investigate the fixed point solution $R_2'(\zeta)^*$ 
for Eq.(\ref{LRpartBetaFunc1LREF}). 
Since Eq.(\ref{LRpartBetaFunc1LREF}) is linear in the function $R_2'(\zeta)$, 
the fixed point equation $\partial_tR_2'(\zeta)^*=0$ can be solved analytically. 
The fixed point equation $\partial_tR_2'(\zeta)^*=0$ 
takes the form 
\begin{eqnarray}
(1-\zeta^2)(R_1'(1)^*+R_2'(1)^*)R_2'''(\zeta)^*
-(N+1)\zeta(R_1'(1)^*+R_2'(1)^*)R_2''(\zeta)^*
+[(N-1)(R_1'(1)^*+R_2'(1)^*)-\epsilon]R_2'(\zeta)^*=0.
\label{FuchsLREF1}\nonumber\\
\end{eqnarray}
\end{widetext}
The solutions of this equation have regular singular points at $\zeta=\pm1$ 
for the interval $-1\le\zeta\le1$. 
Under the condition of $|R_2'(\zeta)^*|<\infty$ on the interval $-1\le\zeta\le1$, 
the solutions of Eq.(\ref{FuchsLREF1}) can be expressed 
in terms of the Gaussian hypergeometric function: 
\begin{eqnarray}
&&R_2'(\zeta)^*\nonumber\\
&&=\left\{
\begin{array}{ll}
\!\!C{}_2F_1(x_1,x_2;y;(1-\zeta)/2)&{\mbox{around $\zeta=1$}}\\
\!\!C'{}_2F_1(x_1,x_2;y;(1+\zeta)/2)&{\mbox{around $\zeta=-1$}}
\end{array}
\right.\!\!,
\end{eqnarray}
where $C$ and $C'$ are constants 
fulfilling the condition $|R_2'(\zeta)^*|<\infty$. 
Here, the generalized hypergeometric function is defined by the following series expansion: 
\begin{eqnarray}
&&{}_mF_n(x_1,x_2,\ldots,x_m;y_1,y_2,\ldots,y_n;z)\nonumber\\
&&\equiv\sum_{k=0}^{\infty}
\frac{(x_1)_k(x_2)_k\cdots(x_m)_k}{(y_1)_k(y_2)_k\cdots(y_n)_k}
\frac{z^k}{k!},\\
&&(x)_k=\Gamma(x+k)/\Gamma(x).
\end{eqnarray}
And, $x_1$, $x_2$ and $y$ are 
\begin{eqnarray}
x_1,x_2
\!&=&\!
\frac{1}{2}\biggl[
N\!\pm\!\sqrt{N^2+4\biggl\{N-1-\frac{\epsilon}{R_1'(1)^*+R_2'(1)^*}\biggr\}}
\biggr],\nonumber\\ \\
y\!&=&\!\frac{N+1}{2}.
\end{eqnarray}
Thus the fixed point solution $R_2'(\zeta)^*$ is an analytic function on $\zeta$. 
Next we examine the renormalization group flow of $R_1'(z)$. 
We assume that the functions $R_1'(z)$ and $R_2'(\zeta)$ take the forms 
given by Eqs.(\ref{initialfunctionLRF1}) and (\ref{initialfunctionLRF2}) with $\alpha>0$. 
To keep $|R_1'(z)|$ and $|R_2'(\zeta)|$ finite, 
the following condition on the function (\ref{initialfunctionLRF1}) is required; 
\begin{eqnarray}
{\mbox{
$\alpha=\frac{1}{2}\quad$or$\quad\alpha\ge1$
}}.
\label{singular condition for LREF}
\end{eqnarray}
The fixed point solution $R_1'(z)^*$ also has the same singularity. 
Only in the case of $\alpha=1/2$, $R''(1)^*$ diverges.

\subsection{Stability of cuspless fixed points}
The critical exponents $\eta$ and $\bar{\eta}$ are expressed 
by use of $R_1'(1)^*$ and $R_2'(1)^*$: 
\begin{eqnarray}
\eta&=&2-\sigma,\label{etaLREF}\\
{\bar{\eta}}&=&(N-1)(R_1'(1)^*+R_2'(1)^*)\nonumber\\
&&-(2\sigma+\rho+\epsilon-4).\label{baretaLREF}
\end{eqnarray}
The critical exponent $\nu$ is determined from the inverse of the maximum eigenvalue 
of the $4\times4$ scaling matrix at the fixed point. 
Then, we find the fixed points by solving 
$\partial_tR_1'(1)^*=0$, $\partial_tR_1''(1)^*=0$, 
$\partial_tR_2'(1)^*=0$, and $\partial_tR_2''(1)^*=0$, 
and study their stability. 

The cuspless fixed points are 
\begin{eqnarray}
&&(R_1'(1)^*,R_2'(1)^*,R_1''(1)^*,R_2''(1)^*)\nonumber\\
&&=\biggl(
\frac{\epsilon+\rho}{N},0,
\frac{(\epsilon+\rho)[N-8+\sqrt{(N-2)(N-18)}]}{2N(N+7)},0
\biggr),\label{LRE FP+}\nonumber\\ 
\end{eqnarray}
\begin{eqnarray}
&&(R_1'(1)^*,R_2'(1)^*,R_1''(1)^*,R_2''(1)^*)\nonumber\\
&&=\biggl(
\frac{\epsilon+\rho}{N},0,
\frac{(\epsilon+\rho)[N-8-\sqrt{(N-2)(N-18)}]}{2N(N+7)},0
\biggr),\label{LRE FP-}\nonumber\\ 
\end{eqnarray}
\begin{eqnarray}
&&(R_1'(1)^*,R_2'(1)^*,R_1''(1)^*,R_2''(1)^*)\nonumber\\
&&=\biggl(
\frac{\epsilon^2}{(N-1)^2\rho},
\frac{\epsilon^2[(N-1){\tilde{\epsilon}}-N]}{(N-1)^2\rho},
\nonumber\\
&&\frac{\epsilon
[(N-1)\tilde{\epsilon}-8
+\sqrt{\{(N-1)\tilde{\epsilon}-8\}^2-4(N+7)}]
}{2(N+7)(N-1)},
0
\biggr),\label{LREF FP+}\nonumber\\ 
\end{eqnarray}
\begin{eqnarray}
&&(R_1'(1)^*,R_2'(1)^*,R_1''(1)^*,R_2''(1)^*)\nonumber\\
&&=\biggl(
\frac{\epsilon^2}{(N-1)^2\rho},
\frac{\epsilon^2[(N-1){\tilde{\epsilon}}-N]}{(N-1)^2\rho},
\nonumber\\
&&\frac{\epsilon
[(N-1)\tilde{\epsilon}-8
-\sqrt{\{(N-1)\tilde{\epsilon}-8\}^2-4(N+7)}]
}{2(N+7)(N-1)},
0
\biggr).\label{LREF FP-}\nonumber\\
\end{eqnarray}
Here we have introduced the reduced variable ${\tilde{\epsilon}}$: 
\begin{eqnarray}
{\tilde{\epsilon}}=\frac{\epsilon+\rho}{\epsilon}. 
\end{eqnarray}
The stability of the cuspless fixed points with respect to the cuspless perturbation 
can be investigated by calculating eigenvalues of the $4\times 4$ scaling matrix 
whose elements are the first derivatives of the beta functions 
$\partial_tR_1'(1)$, $\partial_tR_1''(1)$, 
$\partial_tR_2'(1)$, and $\partial_tR_2''(1)$ 
at the cuspless fixed points. 

The cuspless fixed points (\ref{LRE FP+}) and (\ref{LRE FP-}) exist for $N\ge18$. 
The eigenvalues $\lambda_1,\ldots,\lambda_4$ of the scaling matrix 
at the cuspless fixed points (\ref{LRE FP+}) and (\ref{LRE FP-})
are given by 
\begin{eqnarray}
\lambda_1&=&\epsilon+\rho,\label{me1}\\
\lambda_2&=&-\epsilon\biggl(1-\frac{N-1}{N}{\tilde{\epsilon}}\biggr),
\label{LRE2ev}\\
\lambda_3^{\pm}&=&\pm(\epsilon+\rho)\frac{\sqrt{(N-2)(N-18)}}{N},
\label{LRE2ev2}\\
\lambda_4&=&-\epsilon\biggl(1+\frac{2}{N}{\tilde{\epsilon}}\biggr),
\end{eqnarray}
Thus the fixed point (\ref{LRE FP+}) is multiply unstable. 
If $1\le{\tilde{\epsilon}}<N/(N-1)$ or $0\le\rho<\epsilon/(N-1)$, 
the fixed point (\ref{LRE FP-}) is singly unstable.  
Due to $R_2'(1)^*=0$, the long-range correlations of random fields and random anisotropies 
are irrelevant, 
and thus the fixed point (\ref{LRE FP-}) governs the phase transition 
in the system with LRE. 
Whereas, in $N\le 18$, 
the cuspless fixed points (\ref{LRE FP+}) and (\ref{LRE FP-}) merge and annihilate, 
and thus the beta functions have no cuspless fixed point of O$(\epsilon)$. 

The cuspless fixed points (\ref{LREF FP+}) and (\ref{LREF FP-}) exist for 
$N>1$ and 
\begin{eqnarray}
{\tilde{\epsilon}}\ge{\tilde{\epsilon}}_{\rm cuspless}=\frac{8+2\sqrt{N+7}}{N-1}. 
\end{eqnarray}
The eigenvalues $\lambda_1,\ldots,\lambda_4$ of the scaling matrix 
at the cuspless fixed points (\ref{LREF FP+}) and (\ref{LREF FP-})
are given by 
\begin{eqnarray}
\lambda_1
&=&\epsilon
\biggl[
\frac{N}{N-1}
-\frac{{\tilde{\epsilon}}}{2}
+\frac{{\tilde{\epsilon}}}{2}
\sqrt{1+\frac{4[N-{\tilde{\epsilon}}(N-1)]}{{\tilde{\epsilon}}^2(N-1)^2}}
\biggr],\label{me3}\\
\lambda_2
&=&\epsilon
\biggl[
\frac{N}{N-1}
-\frac{{\tilde{\epsilon}}}{2}
-\frac{{\tilde{\epsilon}}}{2}
\sqrt{1+\frac{4[N-{\tilde{\epsilon}}(N-1)]}{{\tilde{\epsilon}}^2(N-1)^2}}
\biggr],\label{LREF2ev}\\
\lambda_3^{\pm}
&=&\pm(\epsilon+\rho)
\sqrt{1-\frac{4[N-9+4{\tilde{\epsilon}}(N-1)]}{{\tilde{\epsilon}}^2(N-1)^2}},
\label{LREF2ev2}\\
\lambda_4
&=&-\epsilon\frac{N+1}{N-1},
\end{eqnarray}
Thus the fixed point (\ref{LREF FP+}) is multiply unstable. 
If ${\tilde{\epsilon}}>N/(N-1)$ or $\rho>\epsilon/(N-1)$, 
the fixed point (\ref{LREF FP-}) is singly unstable. 
Due to $R_2'(1)^*>0$, 
the effect of the long-range correlation of random fields and random anisotropies 
appears, 
and then the fixed point (\ref{LREF FP-}) governs the phase transition 
in system with LREF. 

\begin{figure}
\begin{center}
\includegraphics[width=80mm]{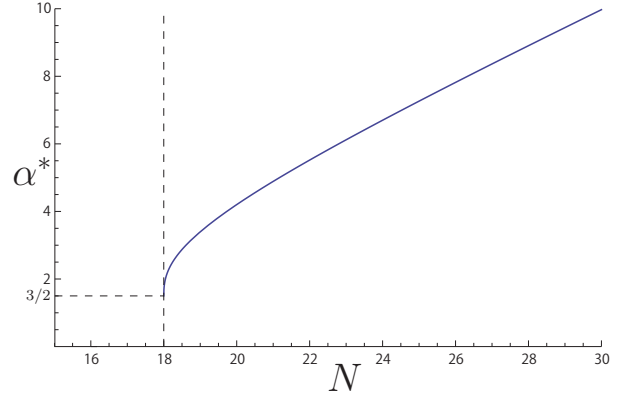}
\end{center}
\caption{
Exponent $\alpha^*=\alpha(N)^*$ characterizing the subcuspy singularity $(1-z)^{\alpha_*}$ 
of the LRE TT FP (\ref{LRE FP-}). 
The ordinate is $\alpha^*$, and the abscissa is $N$. 
$\alpha(N\searrow18)^*=3/2$. 
}
\label{alpha vs N LRE}
\end{figure}

\begin{figure}
\begin{center}
\includegraphics[width=80mm]{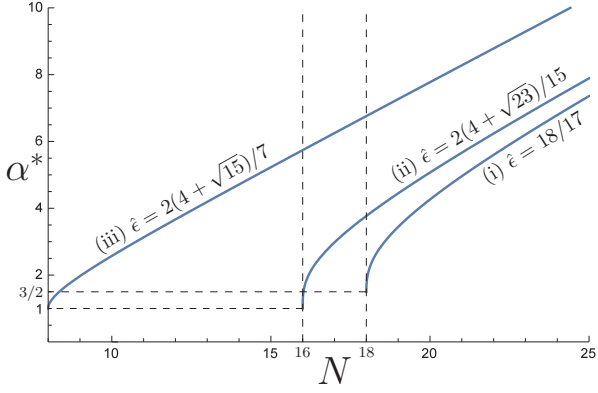}
\end{center}
\caption{
Exponent $\alpha^*=\alpha(N,{\tilde{\epsilon}})^*$ characterizing the subcuspy singularity $(1-z)^{\alpha^*}$ 
of the LREF TT FP (\ref{LREF FP-}) for three values of ${\tilde{\epsilon}}$. 
The ordinate is $\alpha^*$, and the abscissa is $N$. 
(i) For ${\tilde{\epsilon}}=18/17$, the value $N$ of the lower boundary 
above which the LREF TT FP are singly unstable with respect to the cuspless perturbation is $N=18$. 
$\alpha(N\searrow18,{\tilde{\epsilon}})^*=3/2$. 
(ii) For ${\tilde{\epsilon}}=2(4+\sqrt{23})/15$, $N=16$. 
$\alpha(N\searrow16,{\tilde{\epsilon}})^*=1$. 
(iii) For ${\tilde{\epsilon}}=2(4+\sqrt{15})/7$, $N=8$. 
$\alpha(N\searrow8,{\tilde{\epsilon}})^*=1$. 
}
\label{alpha vs N LREF}
\end{figure}

The cuspless fixed points (\ref{LRE FP-}) and (\ref{LREF FP-}) 
have subcuspy singularities $(1-z)^{\alpha_*}$ with a noninteger $\alpha_*\ge1$. 
Then we call the singly unstable fixed points (\ref{LRE FP-}) and (\ref{LREF FP-}) 
as the ``LRE TT FP'' and the ``LREF TT FP'' respectively. 
The power $\alpha_*$ is obtained as follows. 
Calculating the flow of $a_l$ in Eq.(\ref{initialfunctionLRF1}), we have  
\begin{eqnarray}
&&\partial_ta_l=a_l\Lambda_{\alpha+1}(R_1'(1)^*,R_2'(1)^*,R_1''(1)^*,R_2''(1)^*),\\
&&\Lambda_{\alpha+1}(R_1'(1)^*,R_2'(1)^*,R_1''(1)^*,R_2''(1)^*)\nonumber\\
&&=2[R_1'(1)^*+R_2'(1)^*+3(R_1''(1)^*+R_2''(1)^*)]\alpha^2\nonumber\\
&&-[(N-5)(R_1'(1)^*+R_2'(1)^*)\nonumber\\
&&\quad-(N+7)(R_1''(1)^*+R_2''(1)^*)]\alpha\nonumber\\
&&+(N+1)(R_1'(1)^*+R_2'(1)^*+R_1''(1)^*+R_2''(1)^*)\nonumber\\
&&\quad-(\epsilon+\rho).
\label{Eigen for a_k LREF}
\end{eqnarray}
The power $\alpha^*$ is determined from 
\begin{eqnarray}
\Lambda_{\alpha^*+1}(R_1'(1)^*,R_2'(1)^*,R_1''(1)^*,R_2''(1)^*)=0.
\label{subcusp eq. LREF}
\end{eqnarray}
Substituting the LRE TT FP (\ref{LRE FP-}) 
and the LREF TT FP (\ref{LREF FP-}) 
into the above equation, we have explicit expressions for $\alpha^*$, respectively. 
Firstly, substituting the LRE TT FP (\ref{LRE FP-}) 
into Eq.(\ref{subcusp eq. LREF}), 
we obtain the following quadratic equation for $\alpha^*$: 
\begin{eqnarray}
&&\biggl(2+\frac{3}{N+7}[N-8-\sqrt{(N-2)(N-18)}]\biggr){\alpha^*}^2\nonumber\\
&&-\biggl(N-5-\frac{1}{2}[N-8-\sqrt{(N-2)(N-18)}]\biggr)\alpha^*\nonumber\\
&&+1+\frac{N+1}{2(N+7)}[N-8-\sqrt{(N-2)(N-18)}]=0.
\label{LRE qe}
\end{eqnarray}
Solving the above equation, we have the solution $\alpha^*$ as a function of $N$. 
It goes to $N/2+{\rm O}(1)$ at large $N$. 
The solution is the same as that of the system with SR \cite{TT3,TT4}. 
The graph of $\alpha^*=\alpha^*(N)$ is depicted in Fig. \ref{alpha vs N LRE}. 
Next, substituting the LREF TT FP (\ref{LREF FP-}) 
into Eq.(\ref{subcusp eq. LREF}), 
we obtain the following quadratic equation for $\alpha^*$: 
\begin{eqnarray}
&&\biggl(
2+\frac{3}{N+7}[(N-1){\tilde{\epsilon}}-8\nonumber\\
&&\quad-\sqrt{\{(N-1){\tilde{\epsilon}}-8\}^2-4(N+7)}]
\biggr){\alpha^*}^2\nonumber\\
&&-\biggl(
N-5-\frac{1}{2}[(N-1){\tilde{\epsilon}}-8\nonumber\\
&&\quad-\sqrt{\{(N-1){\tilde{\epsilon}}-8\}^2-4(N+7)}]
\biggr)\alpha^*\nonumber\\
&&+N+1-(N-1){\tilde{\epsilon}}
+\frac{N+1}{2(N+7)}[(N-1){\tilde{\epsilon}}-8\nonumber\\
&&\quad-\sqrt{\{(N-1){\tilde{\epsilon}}-8\}^2-4(N+7)}]=0.
\label{LREF qe}
\end{eqnarray}
Solving the above equation, we obtain the solution $\alpha^*$ as a function of $N$ and ${\tilde{\epsilon}}$. 
It goes to $N/2+{\rm O}(1)$ at large $N$. 
The graphs of $\alpha^*=\alpha(N,{\tilde{\epsilon}})^*$ for some values of ${\tilde{\epsilon}}$ 
are depicted in Fig. \ref{alpha vs N LREF}. 

We investigate the stability of the TT fixed points with respect to the cuspy perturbation.  
It can be done by calculating the eigenvalue $\lambda$ 
relating to the cuspy deformation from the TT fixed point, 
which is given by 
\begin{eqnarray}
\lambda
=\Lambda_{3/2}(R_1'(1)^*,R_2'(1)^*,R_1''(1)^*,R_2''(1)^*).
\label{ev wrt cuspy pert LREF}
\end{eqnarray}
Substituting the LRE TT FP (\ref{LRE FP-}) 
and the LREF TT FP (\ref{LREF FP-}) 
into the above equation, we obtain explicit expressions for $\lambda$, respectively. 
The eigenvalues $\lambda_{\rm LRE}$ for the system with LRE
and $\lambda_{\rm LREF}$ for the system with LREF 
are as follows: 
\begin{eqnarray}
\lambda_{\rm LRE}
&=&-\frac{(\epsilon+\rho)(N-2)}{4N(N+7)}\nonumber\\
&&\times\biggl[3(N+4)\sqrt{\frac{N-18}{N-2}}-N+8\biggr].\\
\lambda_{\rm LREF}
&=&-\frac{\epsilon}{4(N-1)(N+7)}\nonumber\\
&&\times[
3(N+4)\sqrt{[(N-1){\tilde{\epsilon}}-8]^2-4(N+7)}\nonumber\\
&&-2(N+7)(N-8)+(N+16)[(N-1){\tilde{\epsilon}}-8]].\nonumber\\
\end{eqnarray}

In the case of the system with LRE, we find that, 
below $N=N_{\rm cusp}=2(4+3\sqrt{3})\simeq18.3923\ldots$, 
the eigenvalue $\lambda_{\rm LRE}$ takes a positive value. 
Thus the cuspy perturbation becomes relevant for $N<N_{\rm cusp}$, 
where the LRE TT FP (\ref{LRE FP-}) is multiply unstable 
with respect to the cuspy perturbation. 
Whereas it remains singly stable 
with respect to the perturbation with and without the cuspy behavior for $N>N_{\rm cusp}$. 

In the case of the system with LREF, 
the eigenvalue $\lambda_{\rm LREF}$ takes a positive value below 
\begin{eqnarray}
{\tilde{\epsilon}}
&=&{\tilde{\epsilon}}_{\rm cusp}\nonumber\\
&=&
\frac{3(N+4)\sqrt{N^2-8N+48}-(N^2-24N-64)}{4(N-1)(N-2)}.
\label{ecusp}\nonumber\\
\end{eqnarray}
Since ${\tilde{\epsilon}}_{\rm cusp}\ge{\tilde{\epsilon}}_{\rm cuspless}$ 
for $4(1+\sqrt{7})\le N \le 2(4+3\sqrt{3})$, 
the LREF TT FP (\ref{LREF FP-}) is destabilized by the cuspy perturbation 
for $4(1+\sqrt{7})\le N \le 2(4+3\sqrt{3})$ and ${\tilde{\epsilon}}<{\tilde{\epsilon}}_{\rm cusp}$. 
Even in this case, a singly unstable cuspy fixed point which governs the phase transition in the system 
is considered to exist 
for $4(1+\sqrt{7})\le N \le 2(4+3\sqrt{3})$ and ${\tilde{\epsilon}}<{\tilde{\epsilon}}_{\rm cusp}$. 

Finally, we calculate the eigenfunction 
which belongs to the eigenvalue (\ref{ev wrt cuspy pert LREF}). 
Solving the eigenvalue equation, 
we obtain two solutions. 
One takes the form of $(1-z)^{\alpha_-(\lambda)}$ with $\alpha_-(\lambda)=1/2$ when $z\to 1$, 
and the other takes the form of $(1-z)^{\alpha_+(\lambda)}$. 
Both solutions individually diverge in $z=-1$. 
The physical eigenfunction is represented as a linear combination 
of two solutions of the eigenvalue equation, 
in which the coefficients should be chosen to eliminate the singularities at $z=-1$. 
The power $\alpha_+(\lambda)$ of the function $(1-z)^{\alpha_+(\lambda)}$ 
can be obtained by imposing 
\begin{eqnarray}
&&\Lambda_{\alpha_++1}(R_1'(1)^*,R_2'(1)^*,R_1''(1)^*,R_2''(1)^*)\nonumber\\
&&=\Lambda_{3/2}(R_1'(1)^*,R_2'(1)^*,R_1''(1)^*,R_2''(1)^*).
\label{power of subcusp in cuspy deform for LREF}
\end{eqnarray}

In the case of the system with LRE, 
substituting the LRE TT FP (\ref{LRE FP-}) into Eq. (\ref{power of subcusp in cuspy deform for LREF}), 
we have 
\begin{eqnarray}
\alpha_+(\lambda_{\rm LRE})=\frac{1}{4}(N-10+\sqrt{(N-2)(N-18)}). 
\end{eqnarray}
For $N\ge18$, $\alpha_+(\lambda_{\rm LRE})$ takes $\alpha_+(\lambda_{\rm LRE})\ge2$. 

In the case of the system with LREF, 
substituting the LREF TT FP (\ref{LREF FP-}) into Eq. (\ref{power of subcusp in cuspy deform for LREF}), 
we have 
\begin{eqnarray}
\alpha_+(\lambda_{\rm LREF})
&=&
\frac{(N-14)(N-1){\tilde{\epsilon}}+N^2-10N+40}{2[3(N-1){\tilde{\epsilon}}+N-8]}\nonumber\\
&&+\frac{(N-2)\sqrt{\{(N-1){\tilde{\epsilon}}-8\}^2-4(N+7)}}{3(N-1){\tilde{\epsilon}}+N-8}. \nonumber\\
\end{eqnarray}
The power $\alpha_+(\lambda_{\rm LREF})$ takes 
$\alpha_+(\lambda_{\rm LREF})\ge 1+\sqrt{3}$ 
for $N\ge N_{\rm cusp}$ and ${\tilde{\epsilon}}\ge N/(N-1)$, 
and $\alpha_+(\lambda_{\rm LREF})\ge 1$ 
for $4(1+\sqrt{7})\le N<N_{\rm cusp}$ and ${\tilde{\epsilon}}\ge{\tilde{\epsilon}}_{\rm cuspless}$. 
However, we should note that, 
for $N<4(1+\sqrt{7})$ and in the region of 
\begin{eqnarray}
{\tilde{\epsilon}}_{\rm cuspless}\le{\tilde{\epsilon}}<\frac{N^3+48N-320}{(N-1)(N-8)(N+16)}, 
\label{alpha<1LREF}
\end{eqnarray}
$\alpha_+(\lambda_{\rm LREF})<1$, 
which is in contradiction with the condition (\ref{singular condition for LREF}). 
Thus, for $N<4(1+\sqrt{7})$ and in the region (\ref{alpha<1LREF}), 
the cuspy deformation from the LREF TT FP (\ref{LREF FP-}) is unphysical. 
Then, the destabilization of the LREF TT FP (\ref{LREF FP-}) by the cuspy perturbation 
does not occur for ${\tilde{\epsilon}}>2(10+21\sqrt{7})/103$. 

\begin{figure}
\begin{center}
\setlength{\unitlength}{1mm}
\begin{picture}(80, 40)(0,0)
     \put(0,-5){ 
\includegraphics[width=72.5mm]{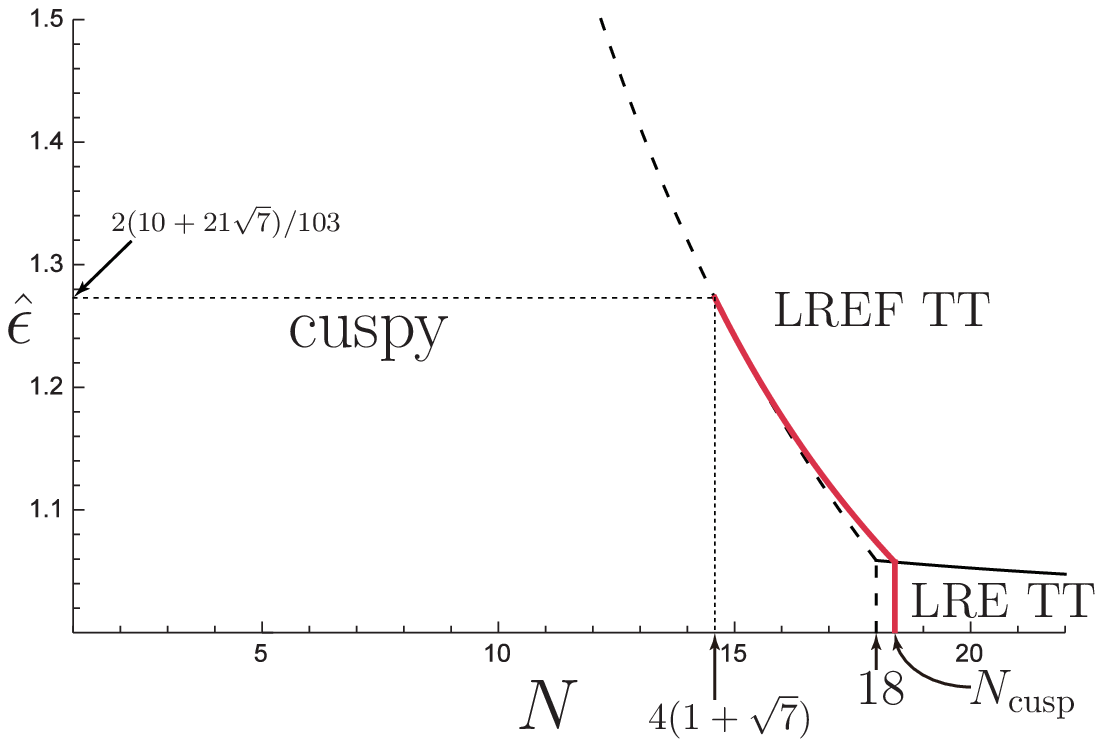}
		}
\end{picture}
\end{center}
\caption{
The regions where various FPs are singly unstable. 
The ordinate is ${\tilde{\epsilon}}(=(\epsilon+\rho)/\epsilon)$, 
and the abscissa is $N$. 
The broken line (black) denotes the lower boundary 
above which the LRE TT and the LREF TT FPs are singly unstable against the cuspless perturbation. 
The border line (black line) between the LRE TT and LREF TT FPs 
is given by ${\hat{\epsilon}}=N/(N-1)$. 
The solid line (red line) denotes the lower boundary 
above which the LRE TT and LREF TT FPs are singly unstable against the cuspy perturbation.
}
\label{phasediagramLREF}
\end{figure}

The regions where the various fixed points are singly unstable are depicted 
in Fig. \ref{phasediagramLREF}. 
Outside the areas where the LRE TT and the LREF TT FPs are singly unstable, 
the cuspy fixed point is considered to control the critical behavior in the system. 
Particularly, in the region of $1\le{\tilde{\epsilon}}<2(10+21\sqrt{7})/103\simeq1.27$, 
the destabilization of the LRE TT and the LREF TT FPs by the cuspy perturbation is caused 
at $N_{\rm cusp}$ for LRE TT FP, and at ${\tilde{\epsilon}}_{\rm cusp}$ for the LREF TT FP, respectively. 

In the following sections, we carefully examine the critical phenomena governed by 
the LRE TT FP (\ref{LRE FP-}) and the LREF TT FP (\ref{LREF FP-}).

\section{Critical phenomena in the system with LRE in $2\sigma+\rho+\epsilon$ dimensions}\label{LRESRF}
In this section we study the critical phenomena controlled by the LRE TT FP (\ref{LRE FP-}). 
We calculate the critical exponents $\eta$, ${\bar{\eta}}$ and $\nu$ at O($\epsilon$). 
We put $\rho=0$, and investigate the $d\to d-\sigma$ dimensional reduction. 
We also discuss the relations between $\eta$ and ${\bar{\eta}}$, 
and present the result for the critical value $\sigma_*$. 

Substituting the LRE TT FP (\ref{LRE FP-}) into Eqs.(\ref{etaLREF}) and (\ref{baretaLREF}), 
we obtain the critical exponents $\eta_{\rm LRE}$ and ${\bar{\eta}}_{\rm LRE}$ 
at the LRE TT fixed point (\ref{LRE FP-}): 
\begin{eqnarray}
\eta_{\rm LRE}&=&2-\sigma,\\
{\bar{\eta}}_{\rm LRE}&=&4-2\sigma-\frac{\epsilon+\rho}{N}. 
\end{eqnarray}
These exponents satisfy $\eta_{\rm LRE}\ge(4-d)/2$, 
${\bar{\eta}}_{\rm LRE}\ge4-d$, and the Schwartz-Soffer inequality 
${\bar{\eta}}_{\rm LRE}\le 2\eta_{\rm LRE}$. 
In the large $N$ limit, the relation 
between $\eta_{\rm LRE}$ and ${\bar{\eta}}_{\rm LRE}$ 
satisfies ${\bar{\eta}}_{\rm LRE}=2\eta_{\rm LRE}$, 
which is identical to the result of the previous study for the critical properties 
of the random field spherical model by Vojta and Schreiber \cite{VS2}. 
For finite $N$ but $N>N_{\rm cusp}$, the relation 
between $\eta_{\rm LRE}$ and ${\bar{\eta}}_{\rm LRE}$ 
satisfies $2\eta_{\rm LRE}-{\bar{\eta}}_{\rm LRE}
=(\epsilon+\rho)/N$ for $\epsilon=d-2\sigma-\rho$. 
Our result is consistent with the result of $1/N$ expansion study by Bray \cite{Br}.  
He showed $2\eta_{\rm LRE}-{\bar{\eta}}_{\rm LRE}=\epsilon/N$
for $\epsilon=d-2\sigma$ by the use of the $1/N$ expansion. 
Thus, the relation $2\eta_{\rm LRE}-{\bar{\eta}}_{\rm LRE}=(d-2\sigma)/N$ 
holds in the region where the scaling behavior in the system is controlled by the LRE TT fixed point. 

We turn to compute the exponent $\nu_{\rm LRE}$ of the correlation length. 
The critical exponent $\nu_{\rm LRE}$ is determined from the inverse of 
the maximal eigenvalue given by Eq.(\ref{me1}). 
Thus, we obtain the inverse of the critical exponent $\nu_{\rm LRE}$ as 
\begin{eqnarray}
\nu_{\rm LRE}^{-1}=\epsilon+\rho. 
\end{eqnarray}

If we put $\rho=0$, the spatial dimension in the present system becomes $d=2\sigma+\epsilon$, 
and then $\nu_{\rm LRE}^{-1}$ is 
\begin{eqnarray}
\nu_{\rm LRE}^{-1}=\epsilon, 
\end{eqnarray}
which is in agreement with that in the pure long-range system in $\sigma$ dimensions less \cite{BZG}.  
Therefore, the $d\to d-\sigma$ dimensional reduction holds at O($\epsilon$), 
and its validity is recognized only for $N>N_{\rm cusp}$. 

The relation between $\eta_{\rm LRE}$ and ${\bar{\eta}}_{\rm LRE}$ is classified on the basis of the value of $\sigma$ as follows: 
\begin{eqnarray}
&&1.\quad\sigma<2-\frac{\epsilon+\rho}{N}:\,{\bar{\eta}}_{\rm LRE}>\eta_{\rm LRE},\label{DRFLRE}\\
&&2.\quad\sigma=2-\frac{\epsilon+\rho}{N}:\,{\bar{\eta}}_{\rm LRE}=\eta_{\rm LRE},\label{DRTLRE}\\
&&3.\quad\sigma>2-\frac{\epsilon+\rho}{N}:\,{\bar{\eta}}_{\rm LRE}<\eta_{\rm LRE},
\end{eqnarray}
for $N>N_{\rm cusp}$. 
Since $\eta_{\rm LRE}\le{\bar{\eta}}_{\rm LRE}\le 2\eta_{\rm LRE}$, 
the case 3 is unphysical. 
Thus, the critical value $\sigma=\sigma_{*}$ 
which separates between the long-range and the short-range exchange regimes of the theory is
\begin{eqnarray}
\sigma_{*}=2-\frac{\epsilon+\rho}{N}. 
\label{LSLRE}
\end{eqnarray}
Here we comment on the critical value $\sigma_{*}$. 
If $\sigma>2-(\epsilon+\rho)/2$, 
the spatial dimension in the present system is above four. 
Then we put $d=2\sigma+\rho+\epsilon=4+\epsilon'$ ($0<\epsilon'\ll1$). 
The critical value $\sigma_{*}$ is rewritten in terms of $\epsilon'$ as follows: 
\begin{eqnarray}
\sigma_{*}=2-\frac{\epsilon'}{N-2}. 
\end{eqnarray}
Since the exponent $\eta$ of the random field O($N$) spin model 
with SR in $4+\epsilon'$ dimensions is $\eta=\eta_{\rm SR}=\epsilon'/(N-2)$ 
at O($\epsilon'$) and for $N>N_{\rm cusp}$, 
our result confirms that the critical value $\sigma_{*}$ 
which separates between the long-range and the short-range exchange regimes of the theory is 
\begin{eqnarray}
\sigma_{*}=2-\eta_{\rm SR}. 
\end{eqnarray}

\section{Critical phenomena in system with LREF in $2\sigma+\rho+\epsilon$ dimensions}\label{LRELRF}
In this section we study the critical phenomena controlled by the LREF TT fixed point (\ref{LREF FP-}). 
We calculate the critical exponents $\eta$, ${\bar{\eta}}$, and $\nu$, 
and investigate the $d\to d-\sigma-\rho$ dimensional reduction and the $d\to d-2$ dimensional reduction.  

\begin{figure}
\begin{center}
\includegraphics[width=85mm]{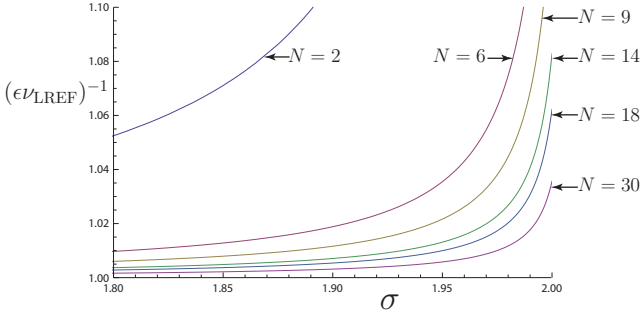}
\end{center}
\caption{
The graphs of Eq.(\ref{LREFnu}) in $\sigma\in[1.80, 2.00]$. 
Here we have put $\rho=2-\sigma$ in Eq.(\ref{LREFnu}). 
Then the reduced variable ${\tilde{\epsilon}}$ takes ${\tilde{\epsilon}}=(\epsilon+2-\sigma)/\epsilon$. 
The ordinate is $(\epsilon\nu_{\rm LREF})^{-1}$, 
and the abscissa is $\sigma$. 
Here, we have set $\epsilon=0.01$. 
Then, ${\tilde{\epsilon}}=(2.01-\sigma)/0.01$. 
}
\label{nu vs s DR}
\end{figure}

Substituting the LREF TT FP (\ref{LREF FP-}) into Eqs. (\ref{etaLREF}) and (\ref{baretaLREF}),  
we obtain the critical exponents $\eta_{\rm LREF}$ and ${\bar{\eta}}_{\rm LREF}$ 
at LREF TT fixed point (\ref{LREF FP-}):  
\begin{eqnarray}
\eta_{\rm LREF}&=&2-\sigma,\\
{\bar{\eta}}_{\rm LREF}&=&4-2\sigma-\rho.
\end{eqnarray}
These exponents satisfy the Schwartz-Soffer inequality 
${\bar{\eta}}_{\rm LREF}\le 2\eta_{\rm LREF}$ \cite{SS},  
and saturate the generalized Schwartz-Soffer inequality 
${\bar{\eta}}_{\rm LREF}\le 2\eta_{\rm LREF}-\rho$ \cite{VS1}.  
And the inverse of the critical exponent $\nu_{\rm LREF}$ is 
\begin{eqnarray}
\nu_{\rm LREF}^{-1}
=\epsilon
\biggl[
\frac{N}{N-1}-\frac{{\tilde{\epsilon}}}{2}
+\frac{{\tilde{\epsilon}}}{2}
\sqrt{1+\frac{4[N-{\tilde{\epsilon}}(N-1)]}{{\tilde{\epsilon}}^2(N-1)^2}}
\biggr].\nonumber\\
\label{LREFnu}
\end{eqnarray}
Since $\lim_{N\to\infty}\nu_{\rm LREF}^{-1}=\epsilon$, in the large $N$ limit, the exponent ${\nu_{\rm LREF}}$ agrees with 
that of the pure long-range system in $\sigma+\rho$ dimensions less \cite{BZG}. 
However, as long as $N$ is finite,  $\nu_{\rm LREF}^{-1}\neq\epsilon$. 
Thus, the $d\to d-\sigma-\rho$ dimensional reduction is broken for finite $N$. 
Hence, the $d\to d-2$ dimensional reduction in the case of $\rho=2-\sigma$ is also broken for finite $N$,  
although the exponents $\eta_{\rm LREF}$ and ${\bar{\eta}}_{\rm LREF}$ satisfy 
${\bar{\eta}}_{\rm LREF}=\eta_{\rm LREF}=2-\sigma$ for $N>1$ and $\sigma>2-\epsilon/(N-1)$. 
The graphs of $(\epsilon\nu_{\rm LREF})^{-1}$ for some values of $N$ 
are depicted in Fig. \ref{nu vs s DR}. 
It shows that $(\epsilon\nu_{\rm LREF})^{-1}$ tends to draw to $1$ 
as the value of the parameter $\sigma$ decreases. 
Then one expects that $(\epsilon\nu_{\rm LREF})^{-1}$ reaches $1$ 
if the value of the parameter $\sigma$ decreases even further. 
However, it is impossible to study within the present framework, 
since the nontrivial fixed point of ${\rm O}(\epsilon)$ disappears. 

\section{Summary}\label{Summary}
In this paper we have reexamined the critical phenomena 
of the long-range random field O($N$) spin model near the lower critical dimension 
by using the ${\rm O}(N)$ nonlinear-sigma model 
with the random fields and all possible higher-rank random anisotropies. 
By the use of the perturbative functional renormalization group, 
we have investigated the stability of the analytic fixed points in the one-loop beta functions. 
Also, we have calculated the critical exponents $\eta$, ${\bar{\eta}}$, and $\nu$ 
at the analytic fixed point controlling the critical behavior in the system. 

We have shown that the analytic fixed point controlling the critical behavior 
in the system with the long-range correlations of the random fields has the subcusp, 
and that it can be destabilized by the cuspy perturbation 
in both cases where the exchange interactions between spins 
are short ranged and long ranged. 

We have studied the critical phenomena in the spin system with LRE. 
We have investigated the $d\to d-\sigma$ dimensional reduction. 
We have found that there exists the once-unstable analytic fixed point 
corresponding to the $d\to d-\sigma$ dimensional reduction 
for $N>N_{\rm cusp}=2(4+3\sqrt{3})\simeq18.3923\cdots$. 
Although it has the subcusp, the weaker nonanalyticity does not change the value of the fixed point. 
Then the critical exponents $\eta_{\rm LRE}$ and ${\bar{\eta}}_{\rm LRE}$ 
evaluated at the once-unstable analytic fixed point 
are 
$\eta_{\rm LRE}=2-\sigma$ and ${\bar{\eta}}_{\rm LRE}=4-2\sigma-(d-2\sigma)/N$, respectively, 
and satisfy the relation $2\eta_{\rm LRE}-{\bar{\eta}}_{\rm LRE}=(d-2\sigma)/N$. 
The inverse of the exponent $\nu_{\rm LRE}$ takes $\nu_{\rm LRE}^{-1}=\epsilon$ 
at O($\epsilon$) in $\epsilon=d-2\sigma$. 
Therefore, we conclude that the $d\to d-\sigma$ dimensional reduction  
at the leading order of the $d-2\sigma$ expansion 
holds only for $N>N_{\rm cusp}$. 
For $N<N_{\rm cusp}$, the nonanalyticity occurring by the appearance of the linear cusp  
breaks down the $d\to d-\sigma$ dimensional reduction. 
This is considered to violate the simple relation between the exponents. 
Thus, one expects that 
the critical scaling behavior in the spin system with LRE 
is described by three independent exponents \cite{TBT}. 
Moreover, we have also obtained the critical value $\sigma_{*}=2-\eta_{\rm SR}$ 
on the basis of the condition $\eta\le{\bar{\eta}}\le2\eta$. 
Thus, our result supports the prediction that the crossover 
between the long-range and the short-range exchange regimes of the theory 
occurs at $\sigma_*=2-\eta_{\rm sr}$ \cite{Br,Sak,HNH,LB,APR,BrePariRi,DTC,HST,BRRZ}. 

We have studied the critical phenomena in the spin system with LREF. 
We have investigated the $d\to d-\sigma-\rho$ dimensional reduction 
and the $d\to d-2$ dimensional reduction. 
We have found the once-unstable analytic fixed point  
controlling the critical behavior. 
Although it has the subcusp, 
the weaker nonanalyticity does not change the value of the fixed point. 
Then the critical exponents $\eta_{\rm LREF}$ and ${\bar{\eta}}_{\rm LREF}$ 
evaluated at the once-unstable analytic fixed point 
are 
$\eta_{\rm LREF}=2-\sigma$ and ${\bar{\eta}}_{\rm LREF}=4-2\sigma-\rho$, respectively, 
and satisfy $2{\bar{\eta}}_{\rm LREF}-\eta_{\rm LREF}=\rho$. 
However, we have shown that the $d\to d-\sigma-\rho$ dimensional reduction 
does not holds within the present analysis, as far as $N$ is finite;  
the exponent $\nu_{\rm LREF}$ does not coincide with 
that of the pure long-range system in $\sigma+\rho$ dimensions less. 
Thus, the $d\to d-2$ dimensional reduction in the case of $\rho=2-\sigma$ 
is also broken for finite $N$. 
The result does not contradict that in our previous study 
for the three-dimensional long-range random field Ising model \cite{BTTS}. 
Since our present study by the use of the perturbative renormalization group 
has been restricted to $\epsilon+\rho=\epsilon+2-\sigma\sim{\rm O}(\epsilon)$, 
only the breakdown of the $d\to d-2$ dimensional reduction has been observed. 
Then, to study the $d\to d-2$ dimensional reduction and its breakdown 
in the $(\sigma+2+\epsilon)$-dimensional long-range 
random field ${\rm O}(N)$ spin model, 
the nonperturbative analysis are needed.  

Finally, we comment on the validity of the $d\to d-\sigma$ dimensional reduction 
in the system with LRE 
near the lower critical dimension and for $N>N_{\rm cusp}$. 
As shown in the previous works by Young \cite{You} and Bray \cite{Br}, 
the value of $\nu_{\rm LRE}^{-1}$ coincides with 
that of the pure long-range system in $\sigma$ dimensions less 
at the leading order in $\epsilon=d_u-d$ near the upper critical dimension $d_u=3\sigma$. 
However, it fails at O($\epsilon^2$). 
Thus, although we have shown that the $d\to d-\sigma$ dimensional reduction holds 
at the leading order in $\epsilon=d-d_l$ near the lower critical dimension $d_l=2\sigma$
and for $N>N_{\rm cusp}$ in the present work, 
there is room for doubt whether it holds beyond one loop, even if $N>N_{\rm cusp}$. 
Further studies by using the higher-loop calculation should shed light on this problem.

\begin{acknowledgements}
The author would like to thank to Matthieu Tissier and Gilles Tarjus 
for discussions in early stage of this work. 
\end{acknowledgements}

\end{document}